\documentclass[aip,jcp,preprint,amsmath,amssymb]{revtex4-2}

\usepackage{graphicx} 
\usepackage{dcolumn}  
\usepackage{bm}       
\usepackage[utf8]{inputenc}
\usepackage[T1]{fontenc}
\usepackage{mathptmx} 
\usepackage{etoolbox}
\usepackage{physics}
\usepackage{stackengine}
\usepackage{xcolor} 
\usepackage{booktabs}
\usepackage{soul}
\usepackage{ulem}

\begin{document}

\title{Laser-induced, blackbody-radiation–assisted rovibrational cooling of symmetric-top molecular ions: NH$_3^+$ and ND$_3^+ $}

\author{Archisman Sinha}
\affiliation{School of Chemical Sciences, Indian Association for the Cultivation of Science, Kolkata 700 032, India}

\author{Brianna R. Heazlewood}
\affiliation{Department of Physics, University of Liverpool, Liverpool L69 7ZX, United Kingdom}

\author{Nabanita Deb}
\email{nabanita.deb@iacs.res.in}
\affiliation{School of Chemical Sciences, Indian Association for the Cultivation of Science, Kolkata 700 032, India}


\begin{abstract}

Quantum-state preparation of molecular ions is a prerequisite for precision spectroscopy and controlled studies of cold ion--molecule dynamics. While such control has been extensively developed for diatomic ions and proposed for linear polyatomic ions, corresponding strategies for symmetric-top molecular ions remain largely unexplored. We present a theoretical investigation of blackbody-radiation (BBR)–assisted rovibrational dynamics and laser cooling in the symmetric-top ions NH$_3^+$ and ND$_3^+$, prepared in specific ro-vibrational states by resonance enhanced multiphoton ionization (REMPI) of the neutral precursor. State-resolved radiative lifetimes and equilibration times are computed, revealing that vibrationally excited states decay rapidly, while the ground-state redistribution is dominated by slow BBR-driven ro-vibrational transitions as pure rotational transitions are forbidden in the non-polar NH$_3^+$ and ND$_3^+$ ions. BBR-assisted laser pumping via the $\nu_2$ umbrella-bending mode efficiently cools rotational levels within fixed $K$ manifolds; however, $\Delta K = 0$ selection rules induce a bottleneck, limiting access to the absolute rovibrational ground state for some initially prepared states. Isotopic substitution to ND$_3^+$ further slows dynamics due to reduced transition dipoles. At room temperature, these cooling schemes yield more than 90\% and 85\% of the population in selected rovibrational states of the NH$_3^+$ and ND$_3^+$ ions, respectively. In contrast, at temperatures below 100~K, BBR-induced redistribution is strongly suppressed for ions initially produced in the rovibrational ground state $\ket{\nu=0, J, K}$, effectively freezing the population for extended storage times. In comparison, vibrationally excited states exhibit lifetimes on the order of a few milliseconds, independent of the temperature of the BBR field. 

\end{abstract}

\maketitle

\section{Introduction}
Ion–molecule chemistry governs the formation and evolution of molecular complexity in a wide range of low-temperature environments, including the interstellar medium (ISM), planetary ionospheres, and laboratory plasmas.~\cite{carrascosa2017imaging} Owing to long-range Coulomb interactions, many ion–molecule reactions proceed without activation barriers and therefore remain efficient even at temperatures of only a few Kelvin.~\cite{bell2009ion} Under such conditions, reaction rates and branching ratios are dictated not by thermal averages but by the populations of specific rovibrational states,\cite{meyer2017ion} rendering internal quantum-state control a central requirement for both accurate modeling and controlled laboratory studies of cold chemistry.~\cite{vogelius2002blackbody, schneider2010all} As a result, cold ion–molecule chemistry investigations are needed to 
resolve discrepancies between astronomical observations and classical astrochemical models.

Despite this intrinsic state sensitivity, much of our present understanding of ion–molecule chemistry is still derived from measurements performed at or near room temperature ($T \approx 300$~K). Such data provide a limited—and often misleading—picture of gas-phase chemistry in cold environments such as the ISM, where temperatures typically lie well below 100~K and thermal equilibrium frequently breaks down.~\cite{Herbst2001, roueff2013molecular} Under the low-temperature and low-pressure conditions prevalent in space, chemical reactivity is governed by the populations of the lowest-lying rovibrational states rather than by thermal averages, and extrapolation of high-temperature rate coefficients becomes unreliable. Quantum-state–specific effects such as tunneling and scattering resonances, which can dominate cold collisions, are also
not captured by conventional thermal models. Accurate descriptions of cold ion–molecule chemistry 
requires experimental access to internally cold, quantum-state–selected ions—capabilities that remain scarce and constitute a major source of uncertainty in current astrochemical and atmospheric models.

These requirements have driven the development of experimental platforms capable of probing ion–molecule collisions under well-controlled, low-temperature conditions, including multipole traps,\cite{gerlich1992inhomogeneous, wester2009radiofrequency} uniform supersonic-flow (CRESU) reactors,\cite{potapov2017uniform, canosa2008gas} merged- or crossed-beam experiments employing supersonic expansions,\cite{hansen2014efficient, Shagam2015, Meerakker2012} and quadrupole radiofrequency (RF) ion traps.~\cite{Willitsch2008} In particular, the quadrupole ion traps enable molecular ions to be sympathetically cooled to millikelvin translational temperatures via interactions with laser-cooled atomic ions such as $\mathrm{Ca}^+$.~\cite{Willitsch2008, staanum2010rotational, Schiller2017} However, the internal degrees of freedom—rotation and vibration—remain largely decoupled from the translational motion.~\cite{staanum2010rotational, schneider2010all} As a consequence, trapped molecular ions typically equilibrate with the ambient blackbody radiation (BBR) field of the vacuum chamber, resulting in broad thermal rovibrational distributions even in otherwise ultracold environments.~\cite{nd_pccp_2012, deb2014laser, hansen2014efficient, Schiller2017, vogelius2002blackbody}

Recent experiments have demonstrated that explicit preparation of internal quantum states can profoundly modify cold chemical dynamics. The combination of Coulomb-crystallized ions with Stark-deflected or cryogenically cooled neutral beams has enabled reactions to be studied with molecules prepared in absolute ground states, selected rotational levels, or specific conformers.~\cite{chang2013specific, kilaj2018observation} Complementary merged-beam experiments have shown that rotational-state preparation can significantly alter long-range interaction potentials, leading to reaction rates that deviate markedly from classical Arrhenius behavior.~\cite{Shagam2015} In addition, studies of isotopic fractionation in cold collisions have revealed that small differences in zero-point energy—directly linked to internal-state populations—can exert a decisive influence on reaction branching ratios.~\cite{tsikritea2021inverse} Together, these results underscore that meaningful measurements in the cold and ultracold regime require explicit control over the internal quantum states of the reacting species.
Furthermore, especially for experiments conducted in ion traps, the quantum states need not only to be prepared, but also maintained over the long experimental time scales of seconds, minutes or even hours. 


Theoretical investigations of laser-based schemes to achieve and preserve quantum-state selectivity in trapped molecular ions are therefore of central importance, yet they become increasingly challenging as molecular complexity grows. While internal-state control is now well established for diatomic molecular ions,\cite{nd_pccp_2012, tong2010sympathetic, vogels2018scattering, staanum2010rotational, schneider2010all, stollenwerk2020cooling, chou2017preparation, haas2019long} its extension to polyatomic ions remains largely unexplored. To date, detailed theoretical treatments of rovibrational cooling have been reported only for a limited number of polyatomic systems, most notably the linear symmetric acetylene cation~\cite{deb2014laser}, C$_2$H$_2^+$. 

The ammonia cation ($\mathrm{NH}_3^+$) provides a particularly compelling system in this context. As a prototypical planar symmetric-top molecule in its $\tilde{X}^2A_2''$ ground electronic state, NH$_3^+$ serves as a benchmark for exploring the structural and dynamical complexity of polyatomic ions, bridging the gap between diatomic systems and other chemically relevant nonlinear polyatomic molecules.~\cite{Herzberg1966} Beyond its fundamental interest, NH$_3^+$ plays a central role in nitrogen chemistry in both the interstellar medium and planetary ionospheres.~\cite{Herbst1973, rednyk2019reaction} Its reactions with molecular hydrogen and other neutral species exhibit a pronounced temperature dependence, with reaction rates and branching ratios that are highly sensitive to the internal rovibrational state of the ion.~\cite{Bohme1992}
For nonlinear symmetric-top ions such as NH$_3^+$, the challenge is compounded by a dense rovibrational level structure and stringent symmetry-imposed selection rules. Although the absence of pure rotational transitions might suggest reduced coupling to BBR, infrared-active vibrational modes instead provide efficient pathways for BBR-driven population redistribution. In NH$_3^+$, the frequency of the out-of-plane umbrella-bending mode ($\nu_2$) lies close to the peak of the 300~K blackbody spectrum, leading to rapid rotational redistribution even in the absence of external perturbations.

In this work, we present the first state-resolved theoretical investigation of laser-induced, BBR-assisted rovibrational cooling in nonlinear symmetric-top molecular ions, focusing on NH$_3^+$ and ND$_3^+$ prepared by REMPI of the neutral precursor and confined in a quadrupole ion trap. Using a comprehensive rate-equation framework that accounts for spontaneous emission, BBR-driven excitation, and BBR-assisted resonant laser pumping, we show that selective driving of the infrared-active $\nu_2$ umbrella-bending mode enables efficient rotational cooling within fixed $K$ manifolds, despite the absence of allowed pure rotational transitions. By exploiting symmetry-imposed selection rules of the $D{_{3h}}$ point group together with near-degenerate rovibrational transitions, we identify experimentally realistic schemes that achieve high rovibrational state purity at both room temperature and under cryogenic conditions (77~K). These results establish a framework for internal-state control of nonlinear symmetric polyatomic ions, opening new avenues for precision spectroscopy and state-resolved studies of cold ion–molecule reaction dynamics.

\section{\label{methods}Energy Levels, Selection Rules, Vibrational Intensities and Computational Methods for Blackbody-Assisted Decay and Cooling}

\subsection{Vibrational Modes of NH$_3^+$ and ND$_3^+$}

\begin{figure*}[htbp]
    \centering
    \includegraphics[width=0.85\linewidth]{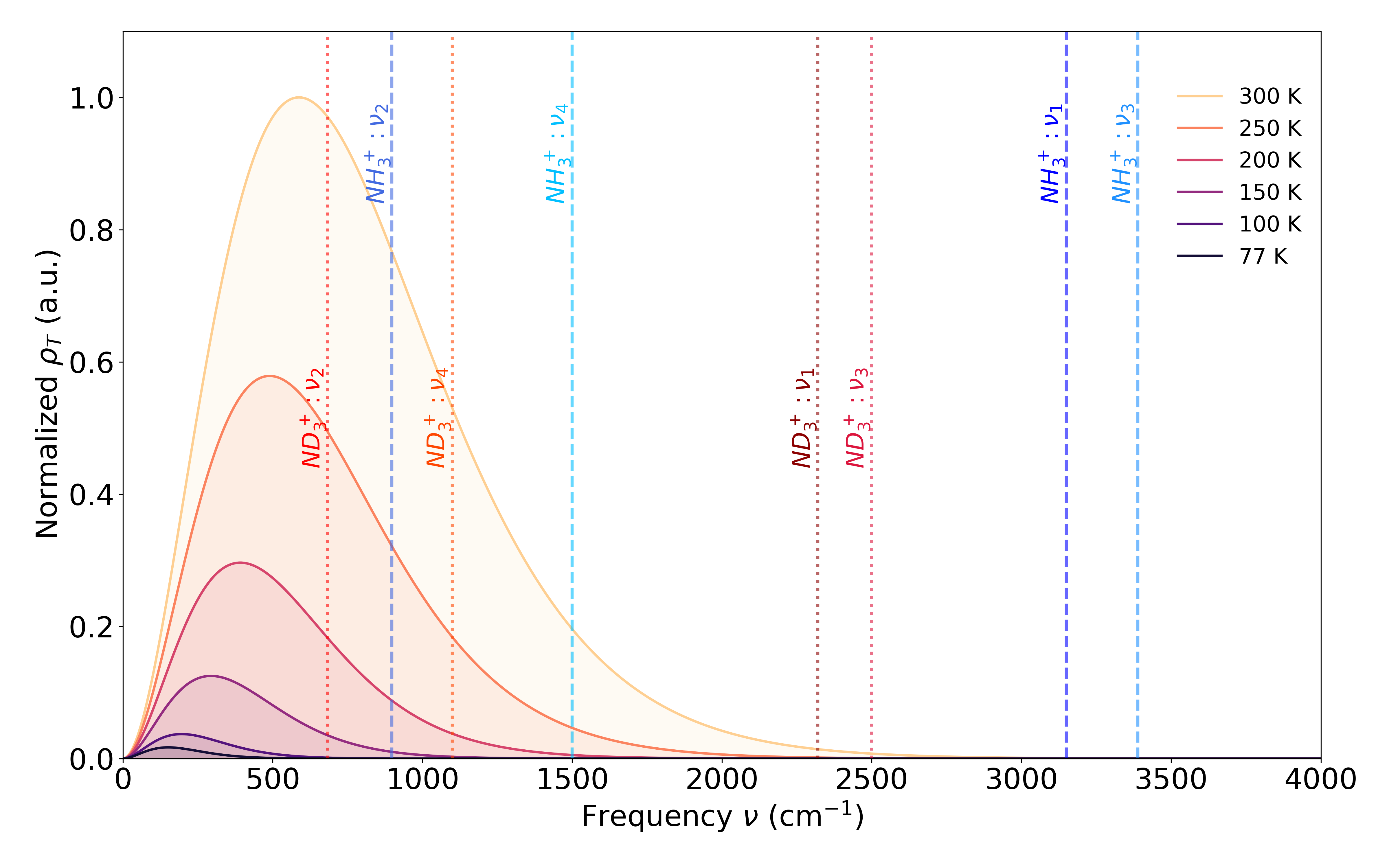}
    \caption{Blackbody radiation (BBR) spectral intensity calculated using Planck’s law for temperatures of 300, 250, 200, 150, 100 and 77~K. The vibrational modes of NH$_3^+$ are indicated by vertical dashed lines in blue and ND$_3^+$ in vertical dotted lines in red indicating their effective spectral overlap with the BBR field.}
    \label{BBR}
\end{figure*}

There are six normal modes of vibration for \( \mathrm{NH_3^+} \): $\nu_1$ (symmetric stretch, 3150 cm$^{-1}$, non-degenerate), $\nu_2$ (umbrella bending mode, 898 cm$^{-1}$, non-degenerate), $\nu_3$ (asymmetric stretch, 3388 cm$^{-1}$, doubly-degenerate) and $\nu_4$ (in-plane scissoring, 1500 cm$^{-1}$, doubly-degenerate)\cite{Herzberg1966}. The only vibrational mode whose frequency overlaps appreciably with the BBR spectrum is $v_2$, the out of plane bending mode (also called the umbrella bending mode). FIG \ref{BBR} shows the BBR intensity, given by Planck's Law (eq.\ref{Planck}), plotted as a function of frequency at different temperatures. From FIG \ref{BBR}, it is clear that the \( \nu_1 \) and \( \nu_3 \) modes exhibit negligible overlap with BBR, even at 300~K. To quantify the spectral characteristics of the background radiation environment, we consider the energy density of a blackbody at temperature $T$. This distribution is fundamentally described by Planck’s law:

\begin{equation}
\rho_T(\omega) = \frac{\hbar \omega^3}{\pi^2 c^3} \cdot \frac{1}{e^{\hbar \omega / k_B T} - 1} .
\label{Planck}
\end{equation}

The relationship defined in Eq. (\ref{Planck}) illustrates how the radiation intensity ($\rho_T$) is distributed across the spectrum ($\omega$). It can be seen that the \( \nu_2 \) mode exhibits a much greater overlap with the BBR field than the \( \nu_4 \) mode at 300~K. However, the overlap for both modes diminishes progressively with decreasing BBR temperature: the involvement of the \( \nu_4 \) mode in thermal equilibration becomes effectively negligible below 200~K, while the \( \nu_2 \) mode becomes negligible below 100~K. The $\nu_2$ bending mode is non-degenerate, corresponding to the \( A_2'' \) irreducible representation \cite{herzberg1956molecularII, Herzberg1966}. 
The vibrational contribution to the total energy of the molecule can be expressed as in a harmonic approximation as:
\begin{equation}
    E_v = G(v_2) = \omega_2(v_2+ \frac{d_2}{2}) 
    \label{eq_for_E}
\end{equation}

\noindent
Here, \( d_2 \) denotes the degeneracy of the vibrational mode, and \( \omega_2 \) is the vibrational frequency of the \( \nu_2 \) mode. In the present study, we restrict our analysis to the first six vibrational levels---namely $\nu_2 = 0, 1, 2, 3, 4, 5$.  
The corresponding vibrational frequencies used in our calculations are summarized in TABLE \ref{tab:constants used}.

Our analysis of potential laser cooling schemes is restricted to the $\nu_2$ mode, as it is the only vibrational mode with an appreciable spectral overlap with the BBR field at 300~K. The $\nu_4$ mode exhibits significantly lower overlap, further compounded by a transition dipole moment that is substantially smaller than that of the $\nu_2$ mode. While one might argue that the red-shifted frequencies in $\text{ND}_3^+$ could increase BBR overlap, this effect is countered by the reduction in the transition dipole moment. For $\text{ND}_3^+$, 
theoretical calculations by Leonard et al. \cite{leonard2001theoretical} indicate that the transition dipole moment for the $\nu_4$ mode is less than 0.1~D, compared to 0.239~D for the $\nu_2$ mode in ND$_3^+$; similar trends are observed for $\text{NH}_3^+$. Since the transition dipole moment is a quadratic term in Einstein's rate equations (see the Methods section), it represents the most sensitive parameter in determining decay dynamics. Consequently, small variations in the transition dipole moment result in relaxation timescales that vary by several orders of magnitude. The combination of poor spectral overlap and an inherently lower transition dipole moment leads to significantly prolonged relaxation times via the $\nu_4$ mode. While the $\nu_4$ mode may become relevant at higher temperatures or over extended timescales, its dynamics are sufficiently slow that the $\nu_2$ mode can be considered independently, as if the $\nu_4$ mode were "frozen" during $\nu_2$ activity. However, the converse is not true: any transition involving the $\nu_4$ mode will invariably be accompanied by $\nu_2$ relaxation due to the latter's much faster decay dynamics. Detailed discussions regarding the $\nu_4$ mode are summarized in Appendices~\ref{app:nu_4 cooling} and \ref{app:leakage}.


\subsection{Rotational Energy Levels of the Out-of-Plane $\nu_2$ Vibrational Mode}

The $\text{NH}_3^+$ molecular ion is an oblate symmetric top. Unlike the pyramidal ($C_{3v}$) structure of neutral ammonia, the cation adopts a planar equilibrium geometry belonging to the $D_{3h}$ point group. This high degree of symmetry results in a vanishing permanent electric dipole moment ($\mu = 0$), which precludes a pure rotational spectrum. Consequently, internal state manipulation must rely on ro-vibrational transitions, such as those involving the umbrella-bending mode ($\nu_2$). 
The rotational energy levels for a given vibrational state $[v]$ are described by the oblate symmetric top expression:\begin{equation}E_{[v]}(J, K) = B_{[v]} J(J+1) - (B_{[v]} - C_{[v]})K^2\label{eq-Fv-JK}\end{equation}\noindent where $B_{[v]}$ and $C_{[v]}$ are the vibrationally averaged rotational constants. \( J \) is the total angular momentum quantum number, with \( K \) the projection of \( J \) onto the molecular axis. To account for the coupling between the $\nu_2$ vibrational mode and the molecular geometry, we utilize the vibrationally averaged constants \cite{lee1991diode}:

\begin{equation}
    \begin{gathered}
        B_{\nu_2} = B_{0} - \alpha_2^B \nu_2 + \gamma_2^B \nu_2 (\nu_2 + 1)    \\
        C_{\nu_2} = C_{0} - \alpha_2^C \nu_2 + \gamma_2^C \nu_2 (\nu_2 + 1) 
    \end{gathered}
    \label{eq-ABC-v}
\end{equation}
\noindent
where $\alpha_2$ and $\gamma_2$ are the vibration-rotation interaction constants. While centrifugal distortion terms ($D_J$, $D_{JK}$, $D_K$) are present in all non-rigid molecules, their contributions are several orders of magnitude smaller than the primary rotational terms and are neglected in the present treatment.


\subsection{Selections Rules of Rovibrational Tranistions within $\nu_2$ mode}

A set of rigorous selection rules are established, governing the allowed transitions between states. As the NH$_3^+$ and ND$_3^+$ isotopologues are planar and lack a permanent dipole moment, pure rotational transitions are forbidden. Transitions can occur only when accompanied by vibrational excitation associated with a non-zero transition dipole moment. The direction of the transition dipole moment determines the nature of the vibrational band. If the dipole moment is aligned along the molecular (symmetry) axis of the molecule, only parallel bands are observed. Conversely, if the dipole moment lies perpendicular to the molecular axis, perpendicular bands occur. In more general cases, where the dipole moment is inclined at an intermediate angle to the figure axis, both parallel and perpendicular bands can appear simultaneously, giving rise to so-called hybrid bands. In our case, the $\nu_2$ vibrational mode 
involves a change in dipole moment along the figure axis. As a result, only parallel bands are allowed, and the selection rule $\Delta K = 0$ applies. Additional selection rules for the rotational quantum number $J$ are as follows:
\begin{align}
\text{For } \Delta K = 0: \quad
&\text{if } K = 0, && \Delta J = \pm 1, \\
&\text{if } K \neq 0, && \Delta J = 0, \pm 1,
\end{align}

\noindent implying that for $K = 0$, only the $P$ and $R$ branches are allowed;
the $Q$ branch is absent. For $K \neq 0$, all three branches can occur. 
It is also important to consider the role of nuclear spin statistics. For NH$_3^+$, in the $K=0$ manifold rotational levels with even $J$ are absent for odd vibrational quantum numbers $\nu$, with the pattern reversed for even $\nu$. In contrast, all rotational levels are allowed in ND$_3^+$, but those with even $J$ carry reduced statistical weights (1:10) for even $\nu$, with the weighting pattern reversed for odd vibrational quantum numbers.

\begin{figure*}[htbp]
    \centering
    {\includegraphics[width=1.00\linewidth, trim=0 0 0 0, clip]{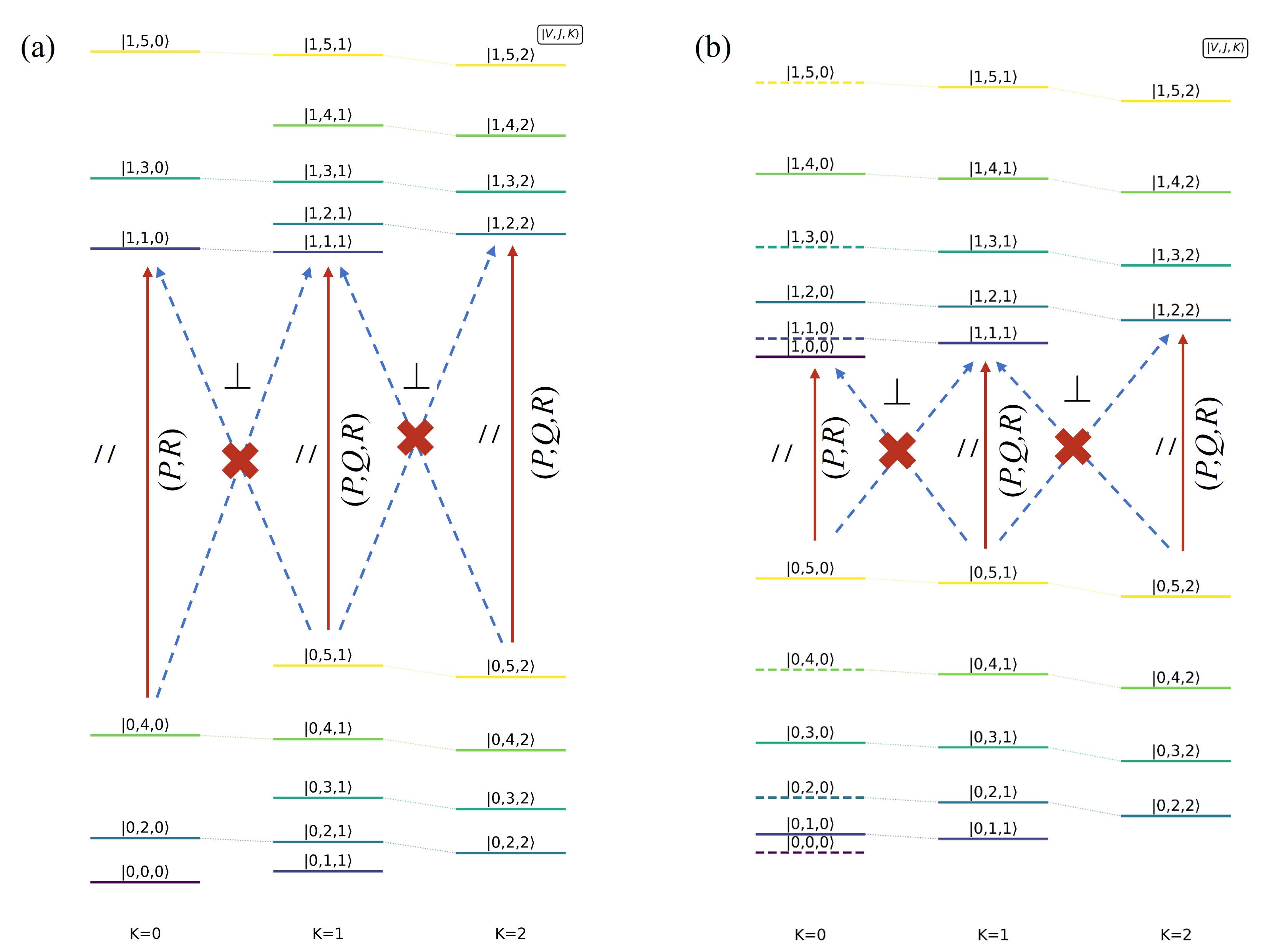}}
    \caption{Energy-level structure illustrating the allowed rovibrational transitions for (a) NH$_3^+$ and (b) ND$_3^+$. For excitation of the $\nu_2$ vibrational mode, only parallel transitions with $\Delta K = 0$ are allowed. Owing to nuclear-spin statistics, the lowest rovibrational level of fermionic NH$_3^+$ is $\ket{0,0,0}$, whereas that of bosonic ND$_3^+$ is $\ket{0,1,0}$. In the $K=0$ manifold of NH$_3^+$, alternate $J$ levels are absent, while in ND$_3^+$ all $J$ levels are present but with modified statistical weights. 
    (The states with lower statistical weight are denoted in dashed -{}- lines). 
    In the $K=1$ manifold, all $J$ levels occur for both isotopologues. For the $K=0$ manifold, only $P$- and $R$-branch transitions are allowed, whereas in the $K=1$ manifold $P$-, $Q$-, and $R$-branch transitions are permitted.}
    \label{Selection rules}
\end{figure*}


\subsection{Rotational and Vibrational Intensities}

From FIG \ref{BBR}, it is evident that even at a temperature of 300~K, the BBR spectrum significantly overlaps with the rovibrational transitions of the molecular ion. The transition moment for rovibrational transitions, $\mu_{\nu,J,K}$, is given as
\begin{equation}
\mu_{(\nu' J' K',\, \nu'' J'' K'')} = \mu_{(\nu', \nu'')} \cdot S_{(J' K',\, J'' K'')},
\end{equation}
\noindent where $S_{(J'K',J''K'')}$ is the rotational line strength of the transition and $\mu_{\nu',\nu''}$ is the vibrational transition dipole moment. 
The rotational line strength for transitions between symmetric top ro-vibrational levels is given by:
\begin{equation}
S(J'K'; J''K'') = (2J'+1)(2J''+1)
\begin{pmatrix}
J'' & 1 & J' \\
K'' & K'-K'' & -K'
\end{pmatrix}^2
\end{equation}

\noindent
This expression originates from the formalism developed by Dennison~\cite{dennison1931infrared,zare1988angular}, 
where double-primed quantum numbers ($J'', K''$) refer to the lower state and single-primed ($J', K'$) to the upper state. Unless otherwise specified, the quantum numbers $J$ and $K$ refer to the lower state, i.e., $J = J''$ and $K = K''$. 

\vspace{0.5em}
\noindent
The rotational line strengths satisfy the following sum rules:
\begin{equation}
\sum_{J', K'} S(J'K'; J''K'') = 2J'' + 1, \qquad 
\sum_{J'', K''} S(J'K'; J''K'') = 2J' + 1.
\end{equation}

\noindent

\noindent
The line strength factors $S_{JK}$ for each branch, derived from the square of the appropriate Wigner 3-j symbols, are given by:
\begin{equation}
    S_{JK} =
    \begin{cases}
        \displaystyle \frac{J^2 - K^2}{J}, & \text{for } \Delta J = -1 \quad (P\text{ branch}) \\
        \displaystyle \frac{(2J+1)K^2}{J(J+1)}, & \text{for } \Delta J = 0 \quad (Q\text{ branch}) \\
        \displaystyle \frac{(J+1)^2 - K^2}{J+1}, & \text{for } \Delta J = +1 \quad (R\text{ branch})
    \end{cases}
    \label{S-JK}
\end{equation}

\begin{figure}[htbp]
    \centering
    \includegraphics[width=0.8\linewidth]{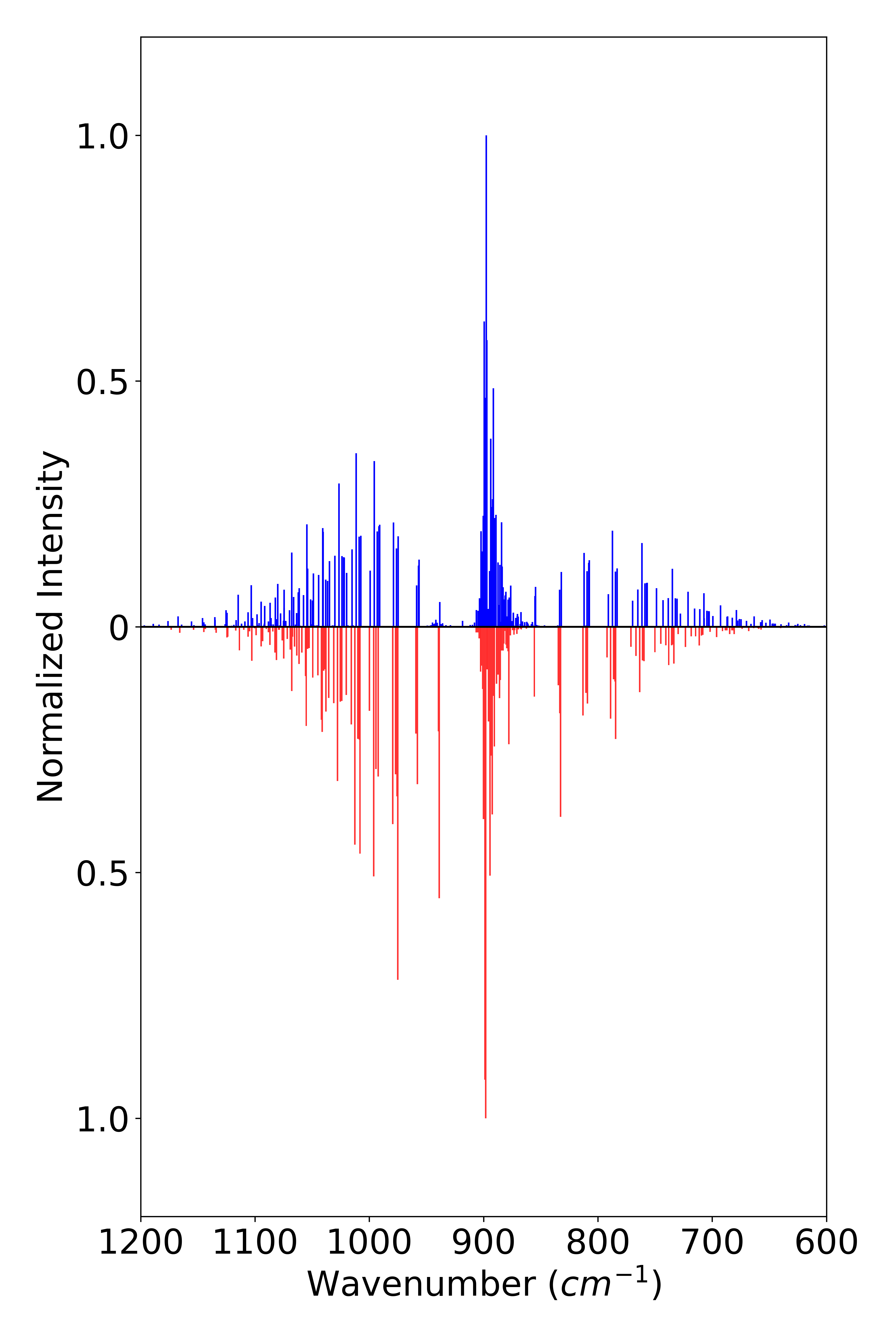}
    \caption{Spectra of NH$_3^+$ involving the $\nu_2$ mode of vibration. The stick spectrum of ro-vibrational transitions as calculated in this work is shown in blue at 300~K. Inverted below this simulation is the computational spectrum (in red) reported by Yurchenko et al\cite{yurchenko2008ab}.}
    \label{Sepctra_comparison}
\end{figure}

To verify the energy levels and validate the rotational line strength factors, we perform spectral simulations using the formulation from Herzberg~\cite{Herzberg1966}. Although derived from a different framework than the one utilized by Yurchenko et al.~\cite{yurchenko2008ab}, the intensity distribution in the infrared band can be consistently expressed as:

\begin{equation}
    I_{JK} = C\nu S_{JK} g_{JK} e^{-E(J,K)/kT}
\end{equation}

\noindent where $S_{JK}$ are the Hönl-London factors, $g_{JK}$ is the statistical weight of the lower state, $\nu$ is the wavenumber of the transition, $k$ is the Boltzmann constant (in cm$^{-1}$) and $C$ is the scaling factor. The good agreement between the simulated spectrum generated in this way and the computed spectrum reported by Yurchenko et al.~\cite{yurchenko2008ab} (see FIG \ref{Sepctra_comparison}) confirms that the spectroscopic data used in the rate equation model are accurate.


\subsection{Computational Methods: Modeling Blackbody-Assisted Decay Rates}

The interaction of BBR with the rotational and vibrational degrees of freedom of a molecule can be quantitatively described using rate equations governing the population dynamics. Specifically, the time evolution of the number density \( n_i \) in a given molecular level \( i\) (i.e., the lower level $e_i<e_j$) is determined by the following BBR rate equation: \cite{nd_pccp_2012}
\begin{equation}
-\frac{dn_i}{dt} = \sum_j B_{ij} \rho_T(\omega_{ij}) n_i - \sum_j \left[ B_{ji} \rho_T(\omega_{ij}) + A_{ji} \right] n_j
\end{equation}
\noindent where $\rho$ is the spectral density at the frequency $\omega$, given (in the absence of additional light sources) by Planck’s radiation law for the temperature of the surroundings.

The Einstein coefficients for spontaneous emission ($A$) and stimulated processes ($B$) are derived from established theory and defined as follows 
\begin{equation}
A_{v'J'K',v''J''K''} = \frac{\omega^3}{3\pi\epsilon_0{\hbar}c^3} \frac{S_{J'K',J''K''}}{2J'+1}\mu_{v'v''}^2 ,
\end{equation}
\begin{equation}
B_{v'J'K',v''J''K''} = \frac{1}{6\epsilon_0\hbar^2} \frac{S_{K'J',J''K''}}{2J'+1}\mu_{v'v''}^2 ,
\end{equation}
\begin{equation}
B_{v''J''K'',v'J'K'} = \frac{2J'+1}{2J''+1} B_{v'J'K',v''J''K''},
\end{equation}
where $\omega$ is the energy difference between the corresponding states and $S_{J'K',J''K''}$ are the Hönl-London Factors~\cite{honl1925intensitaten,dennison1926rotation,reiche1926quantelung,zare1988angular}. The values used in this work are provided in TABLE \ref{tab:constants used}.

\begin{table}[!h]
    \centering
    \renewcommand{\arraystretch}{1.25}  
    \setlength{\tabcolsep}{15pt}  
    \caption{\label{tab:constants used}The parameters used for calculations reported in this work~\cite{maier1989ion,lee1991diode,leonard2001theoretical}.}
    \begin{tabular}{l l l} \hline \hline
        Constants & NH$_3^+$ & ND$_3^+$\\ \hline \hline
        $\omega_2 (1 \xleftarrow{} 0)$ & 898 $cm^{-1}$      & 683 $cm^{-1}$\\ 
        $\omega_2 (2 \xleftarrow{} 1)$ & 941 $cm^{-1}$      & 709 $cm^{-1}$\\
        $\omega_2 (3 \xleftarrow{} 2)$ & 973 $cm^{-1}$      & 729 $cm^{-1}$\\
        $\omega_2 (4 \xleftarrow{} 3)$ & 1000 $cm^{-1}$     & 747 $cm^{-1}$\\
        $\omega_2 (5 \xleftarrow{} 4)$ & 1023 $cm^{-1}$     & 761 $cm^{-1}$\\
        $\omega_2 (6 \xleftarrow{} 5)$ & 1042 $cm^{-1}$     & 774 $cm^{-1}$\\
        $B_{e}$                        & 10.6442 $cm^{-1}$  & 5.3179 $cm^{-1}$\\
        $C_{e}$                        & 5.2476 $cm^{-1}$   & 2.6590 $cm^{-1}$ \\
        $\alpha_B$   & 0.470   & -- \\
        $\gamma_B$   & 0.0110  & -- \\
        $\alpha_C$   & -0.041  & -- \\
        $\gamma_C$   & 0.013   & -- \\
    \hline 
    \end{tabular}
\end{table}


To incorporate the contribution of an excitation laser into the rate equations, the laser intensity was modeled using a Gaussian profile over a narrow spectral bandwidth (3~cm$^{-1}$). This contribution is treated as a sequential addition to the BBR field. When the laser is inactive, we assume that only the BBR contributes to the background radiation:
\begin{equation}
    \rho_{T} = \rho_{\text{Planck}}
\end{equation}
When the laser source is activated, the total spectral density becomes:
\begin{equation}
    \rho_{T} = \rho_{\text{Planck}} + \rho_{\text{laser}}
\end{equation}
This approach is computationally efficient as it avoids the necessity of redefining the Einstein coefficients; instead, the numerical model explicitly accounts for the increased spectral density within the specified frequency range. When a particular transition falls within the frequency window of the laser, the transition undergoes saturation. This saturation occurs for both absorption and the stimulated emission components of the rate equations. Consequently, the population of a specific state is pumped, while the excited state experiences a corresponding increase in stimulated radiation. Thus, the system dynamically achieves a new steady-state equilibrium where the state populations are altered from their original thermal equilibrium values. The spontaneous emission rate remains constant as it does not interact with the radiation field, ensuring its contribution to the system dynamics remains intact.

\section{Results and Discussion}
\label{sec:results}

To investigate the behavior of NH$_3^+$ and ND$_3^+$ molecular ions in a BBR field and to subsequently assess the efficiency of rotational cooling, the precursor ions are assumed to be prepared either in the vibrational ground state ($v=0$) or in the fifth excited level of the $\nu_2$ mode ($\nu_2=5$). These choices are motivated by established experimental protocols. In particular, Zare and co-workers demonstrated that $(2+1)$ REMPI of neutral ammonia via the $B'$ or $C'$ Rydberg states enables the efficient production of NH$_3^+$ ions with high vibrational specificity~\cite{conaway1985vibrational}. Although this scheme can populate vibrational levels ranging from $v=0$ to $v=10$, the $\nu_2=5$ state is selected here as a representative intermediate level that is routinely accessed experimentally. This choice provides sufficient internal energy to probe non-trivial radiative cascade dynamics while remaining experimentally accessible with high state purity.

\subsection{Radiative Lifetime and Equilibration Time of Several $\ket{\nu,J,K}$ States}


{\renewcommand{\arraystretch}{1.25} 
\begin{table}[!h]
\caption{\label{tab:nh3_nd3_lifetime_comparison}Comparison of radiative lifetimes for NH$_3^+$ and ND$_3^+$ quantum states at 300~K and 77~K. The $\nu$ in the state index  denotes vibrational quantum number of the $\nu_2$ bending mode. Note that certain states are missing due to nuclear spin statistics.}
\begin{ruledtabular}
\begin{tabular}{l c c c c}
\multicolumn{5}{c}{ Lifetime}\\
\hline 
 & \multicolumn{2}{c}{at 300~K} & \multicolumn{2}{c}{at 77~K} \\
 \cline{2-3} \cline{4-5}
 $\ket{v, J, K}$ & NH$_3^+$ & ND$_3^+$ & NH$_3^+$ & ND$_3^+$ \\
 \hline
 \noalign{\medskip}
 $\ket{0, 0, 0}$ & 3.2 s & -- & $\gg 3600$ s &  -- \\
 $\ket{0, 1, 0}$ & -- & 4.5 s & -- &  $\gg 3600$ s \\
 $\ket{0, 1, 1}$ & 3.2 s & 4.5 s & $\gg 3600$ s & $\gg 3600$ s  \\
 $\ket{5, 0, 0}$ & -- & 30.2 ms & -- & 32.0 ms  \\
 $\ket{5, 1, 0}$ & 7.4 ms & -- & 7.5 ms & -- \\
 $\ket{5, 1, 1}$ & 7.4 ms & 30.1 ms & 7.5 ms & 31.9 ms  \\
 \noalign{\medskip}
 
 \hline
 \hline
\multicolumn{5}{c}{Equilibration time}\\
\hline 

& \multicolumn{2}{c}{at 300~K} & \multicolumn{2}{c}{at 77~K} \\
 \cline{2-3} \cline{4-5}
 $\ket{v, J, K}$ & NH$_3^+$ & ND$_3^+$ & NH$_3^+$ & ND$_3^+$ \\
 \hline
 \noalign{\medskip}
 $\ket{0, 0, 0}$ & 91 s & -- & $\gg 3600$ s &  -- \\
 $\ket{0, 1, 0}$ & -- & 172 s & -- & $\gg 3600$ s  \\
 $\ket{0, 1, 1}$ & 80 s & 152 s & $\gg 3600$ s & $\gg 3600$ s  \\
 $\ket{5, 0, 0}$ & -- & 172 s & -- & 2 s  \\
 $\ket{5, 1, 0}$ & 80 s & -- & 0.5 s & -- \\
 $\ket{5, 1, 1}$ & 70 s & 133 s & 0.4 s & 1.8 s  \\
\end{tabular}
\end{ruledtabular}
\end{table}
}

The feasibility of rotational cooling in NH$_3^+$ and ND$_3^+$ molecular ions is assessed through a detailed characterization of the spectroscopic properties of the relevant quantum states. The efficiency of vibrational cooling is governed by the competition between spontaneous decay, quantified by the Einstein $A$ coefficients, and thermally induced excitation driven by the ambient BBR field. Consequently, reliable methods for calculating time-dependent population dynamics under varying temperature conditions and for different initial state preparations are of considerable importance.

Following the REMPI schemes established by Zare and coworkers~\cite{conaway1985vibrational}, vibrationally excited NH$_3^+$ molecular ions can be prepared in the $\nu_2=5$ and in the ground $\nu_2 = 0$ levels. Here, the radiative and equilibration lifetimes of the following states are investigated: $\ket{0, 0, 0}$, $\ket{0, 1, 1}$, $\ket{5, 1, 0}$ and $\ket{5, 1, 1}$ for NH$_3^+$; $\ket{0, 1, 0}$, $\ket{0, 1, 1}$, $\ket{5, 0, 0}$ and $\ket{5, 1, 1}$ for ND$_3^+$.

The radiative lifetimes, $t_{1/2} = \left[\sum_j \left( B_{ji}\,\rho_T(\omega) + A_{ji} \right)\right]^{-1},$
are calculated ($i$, $j$ being initial and final states) for the ground states and relevant excited vibrational states of both isotopologues, NH$_3^+$ and ND$_3^+$, and the results are summarized in Table~\ref{tab:nh3_nd3_lifetime_comparison}. Consistent with the vibrational frequency scaling relation $\nu \propto \mu^{-1/2}$, the heavier isotopologue ND$_3^+$ exhibits reduced vibrational level spacings. While this red shift nominally increases the spectral overlap between the vibrational transitions and the BBR field, additional competing effects ultimately govern the overall decay rates. In particular, the transition dipole moments of ND$_3^+$ are significantly smaller~\cite{leonard2001theoretical} than those of NH$_3^+$. 
Because the transition dipole moment enters quadratically into the rate expressions, it represents the most sensitive parameter influencing the radiative lifetimes, such that even modest variations can lead to changes of several orders of magnitude. The resulting decay rates therefore reflect a balance between enhanced BBR coupling due to reduced transition frequencies and suppressed spontaneous emission arising from weaker transition dipole moments.

It is observed that the vibrationally excited states ($\ket{\nu_2 = 5, J, K}$) of ND$_3^+$ persist for approximately four times longer than the corresponding $\ket{\nu_2 = 5, J, K}$ states of NH$_3^+$. This trend is seen at 300~K and 77~K. Nevertheless, in both isotopologues the lifetimes remain on the order of milliseconds and are largely independent of the temperature of the BBR field. 

In contrast, the vibrational ground state ($\ket{\nu = 0, J, K}$) exhibits significantly longer lifetimes of approximately $3.2~\mathrm{s}$ for NH$_3^+$ and $4.5~\mathrm{s}$ for ND$_3^+$ at $300~\mathrm{K}$. However, even these extended lifetimes are insufficient to enable meaningful experimental interrogation of state-selected reaction rates in an ion trap operated at room temperature. 
Interestingly, when the BBR temperature is reduced below $100~\mathrm{K}$, the populations of the $\ket{\nu=0, J, K}$ rotational states 
remain effectively frozen over experimentally relevant timescales. This suppression of blackbody-induced redistribution will enable state-selected ion--molecule reaction studies and precision measurements to be conducted in cryogenic ion traps.

The equilibration time is defined as the point at which the population of every quantum state varies by no more than $0.01\%$ per second, indicating that the system has reached a steady state. The simulations show that, at $300~\mathrm{K}$, ND$_3^+$ requires substantially longer times—approximately a factor of two—to reach thermal equilibrium when prepared in a state-selective manner compared to NH$_3^+$, for both the vibrational ground state and the excited $\nu_2 = 5$ level (see Table~\ref{tab:nh3_nd3_lifetime_comparison}). 

At $77~\mathrm{K}$, ions prepared state-selectively in the vibrationally excited $\ket{\nu_2 = 5, J, K}$ states require approximately $1~\mathrm{s}$ for NH$_3^+$ and less than $5~\mathrm{s}$ for ND$_3^+$ to attain steady state. In contrast, the equilibration of the $\ket{\nu = 0, J, K}$ states is effectively frozen, and the system requires exceedingly long times to reach thermal equilibrium due to the weak spectral overlap with the blackbody radiation field. 

\begin{figure*}[htbp]
    \centering
    \begin{minipage}{0.49\linewidth}
        \centering
        \stackinset{l}{2pt}{t}{2pt}{\textbf{(a)}}{%
            \includegraphics[width=\linewidth, trim={6cm 2cm 5cm 3cm}, clip]{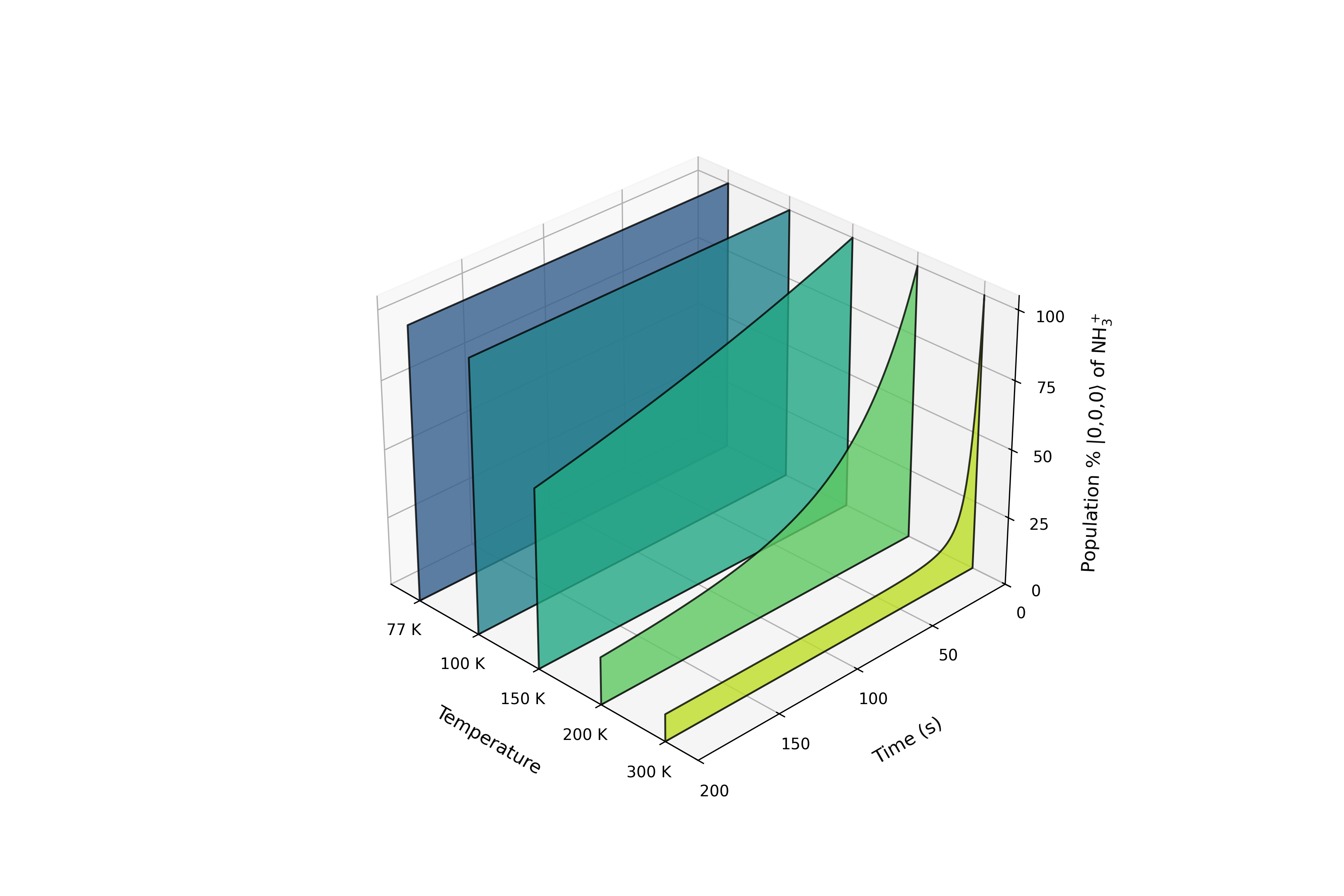}%
        }
    \end{minipage}
    \hfill 
    \begin{minipage}{0.49\linewidth}
        \centering
        \stackinset{l}{2pt}{t}{2pt}{\textbf{(b)}}{%
            \includegraphics[width=\linewidth, trim={6cm 2cm 5cm 3cm}, clip]{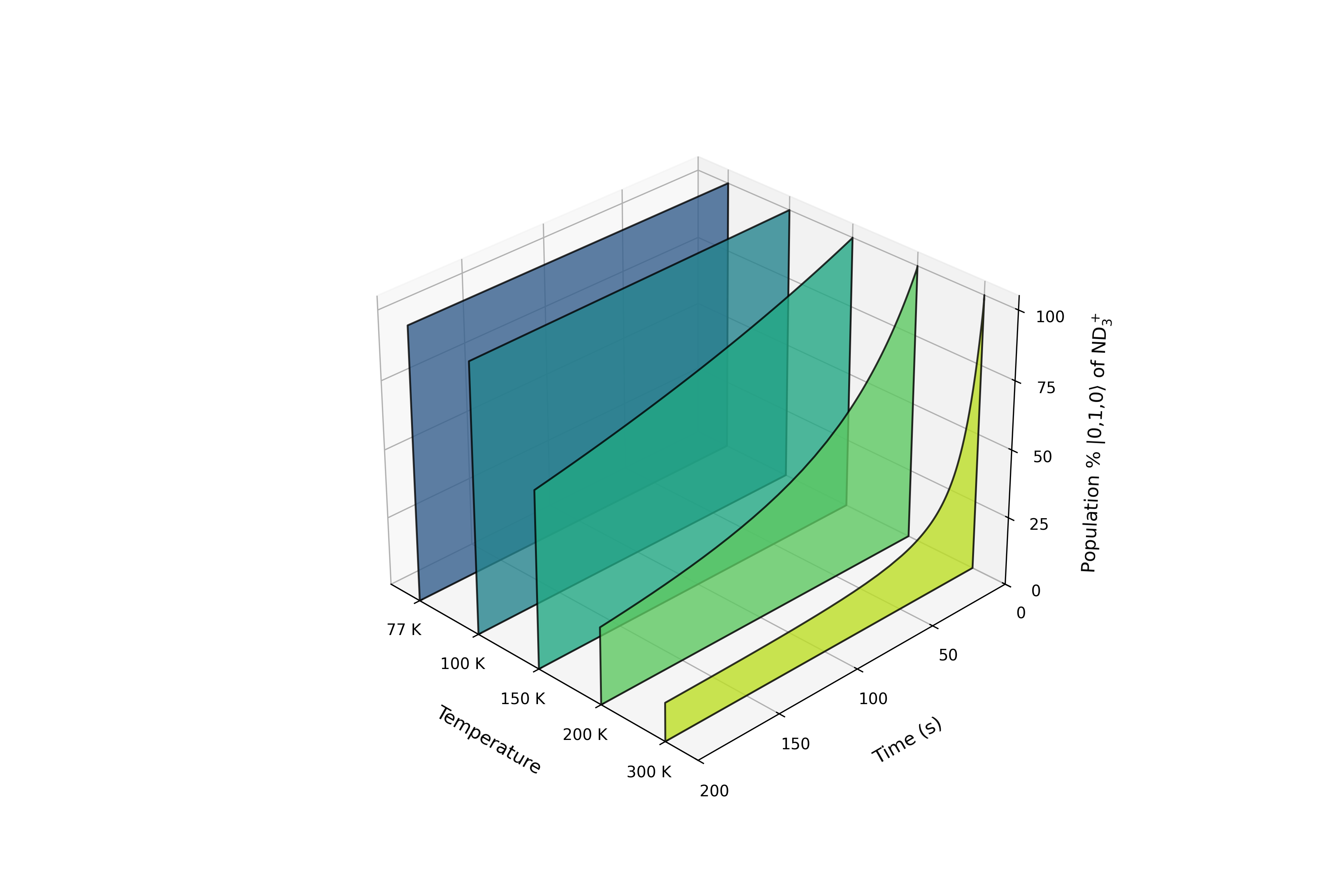}%
        }
    \end{minipage}

    \caption{BBR temperature dependent decay of ground state $|0,0,0\rangle$ NH$_3^+$ (a) and $|0,1,0\rangle$ ND$_3^+$ (b) molecular ions. The decay is slower in the case of ND$_3^+$ because of the reduced value of the transition dipole moment. Nonetheless, the decay freezes at a temperature below 100~K, for both species. The state index is denoted as $\ket{\nu_2,J,K}$}.
    \label{fig:decay_comparison}
\end{figure*}

FIG. \ref{fig:decay_comparison} (a) and (b) illustrates the BBR temperature–dependent decay dynamics of the 
$\ket{0,0,0}$ (NH$_3^+$) and $\ket{0,1,0}$ (ND$_3^+$) states 
at 300~K, 200~K, 150~K, 100~K and 77~K. The decay is comparatively slower for ND$_3^+$ due to its smaller transition dipole moments. Nevertheless, for both isotopologues, the decay is effectively quenched at temperatures below 100~K.

Notably, once the $\ket{\nu_2 = 5, J, K}$ states are populated, they decay rapidly to lower-lying vibrational levels, predominantly via spontaneous emission. As a result, reducing the ambient BBR temperature has a negligible impact on the lifetimes of these vibrationally excited states. 
In contrast, when molecular ions are initially prepared in the vibrational ground state, no energetically lower-lying states are available for decay. Population redistribution can therefore occur only via BBR-induced excitation to higher vibrational levels, followed by subsequent spontaneous emission. This indirect pathway governs the redistribution dynamics and proceeds on comparable timescales for both the $K=0$ and $K=1$ rotational manifolds. At room temperature (300~K), the rapid spontaneous decay of the vibrationally excited states ensures that the overall equilibration times are similar whether the ions are prepared in excited or ground vibrational states. The resulting equilibrium quantum state population distributions are provided in  TABLE~\ref{tabled:equilibrium populations}.

\begin{table}[h!]
    \centering
    \setlength{\tabcolsep}{5pt}
    \caption{\label{tabled:equilibrium populations}Equilibrium population distribution for several states of NH$_3^+$ and ND$_3^+$. In each $K$ manifold the ions are first produced exclusively in the lowest energy level before equilibrating with the BBR field at 300~K.}
    \label{tab:equilibrium_populations}
    \begin{tabular}{lcc @{\hspace{1em}}|@{\hspace{1em}} lcc}
        \toprule
        \multicolumn{3}{c}{\textbf{$K=0$ manifold}} & \multicolumn{3}{c}{\textbf{$K=1$ manifold}} \\
        \cmidrule(l){1-3} \cmidrule(r){4-6}
        \multicolumn{6}{c}{Population (\%)}\\
        State & NH$_3^+$ & ND$_3^+$ &  State & NH$_3^+$ & ND$_3^+$ \\
        \midrule
        $|0,0,0\rangle$  & 9.9   & --    & $|0,1,1\rangle$  & 14.1 & 7.1 \\
        $|0,1,0\rangle$  & --    & 13.9  & $|0,2,1\rangle$  & 19.2 & 10.7 \\
        $|0,2,0\rangle$  & 36.4  & --    & $|0,3,1\rangle$  & 19.8 & 12.9 \\
        $|0,3,0\rangle$  & --    & 25.1 & $|0,4,1\rangle$  & 16.9 & 13.5 \\
        $|0,4,0\rangle$  & 32.1 & --    & $|0,5,1\rangle$  & 12.4 & 12.8 \\
        $|0,5,0\rangle$  & --    & 24.9 & $|0,6,1\rangle$  & 7.9  & 11.1 \\
        $|0,6,0\rangle$  & 15.1 & --    & $|0,7,1\rangle$  & 4.5  & 9.0 \\
        $|0,7,0\rangle$  & --    & 17.5 & $|0,8,1\rangle$  & 2.2  & 6.7 \\
        $|0,8,0\rangle$  & 4.2  & --    & $|0,9,1\rangle$  & 1.0   & 4.8 \\
        $|0,9,0\rangle$  & --    & 9.3  & $|0,10,1\rangle$ & 0.4   & 3.1 \\
        $|0,10,0\rangle$ & 0.9 & --    & $|0,11,1\rangle$ & 0.1  & 2.0 \\
        $|0,11,0\rangle$ & --    & 3.8  & $|0,12,1\rangle$ & <0.1  & 1.1 \\
        $|0,12,0\rangle$ & 0.1  & --    &  \\
        $|0,13,0\rangle$ & --    & 1.0   & \\
        \midrule 
        $|1,0,0\rangle$  & --    & 0.18  &  $|1,1,1\rangle$ & 0.19  & 0.27 \\
        $|1,1,0\rangle$  & 0.4   & -- & $|1,2,1\rangle$ & 0.26  & 0.40 \\
        $|1,2,0\rangle$  & --    & 0.79  & $|1,3,1\rangle$ & 0.27  & 0.48 \\
        \bottomrule
    \end{tabular}
\end{table}

When ions are prepared in the $\ket{\nu=0, J, K}$ states, lowering the ambient BBR temperature has a dramatic impact on both radiative lifetimes and equilibration dynamics. As the temperature decreases, the Planck radiation spectrum shifts to lower wavenumbers and its total spectral radiance is strongly reduced, leading to a substantial decrease in spectral overlap with the molecular vibrational transitions. Consequently, BBR-driven excitation becomes highly inefficient, effectively suppressing population transfer to higher-lying states and arresting internal-state redistribution. Operating in this low temperature regime enables the preparation and long-term storage of molecular ion ensembles with near-unity internal-state purity, providing an attractive platform for precision spectroscopy and controlled ion--molecule reaction studies in the vibrational ground state.


\subsection{BBR-Assisted Rotational Cooling schemes for NH$_3^+$ and ND$_3^+$ molecular ions at 300 K}

\begin{figure}[htbp]
    \centering
    \includegraphics[width=1.0\linewidth]{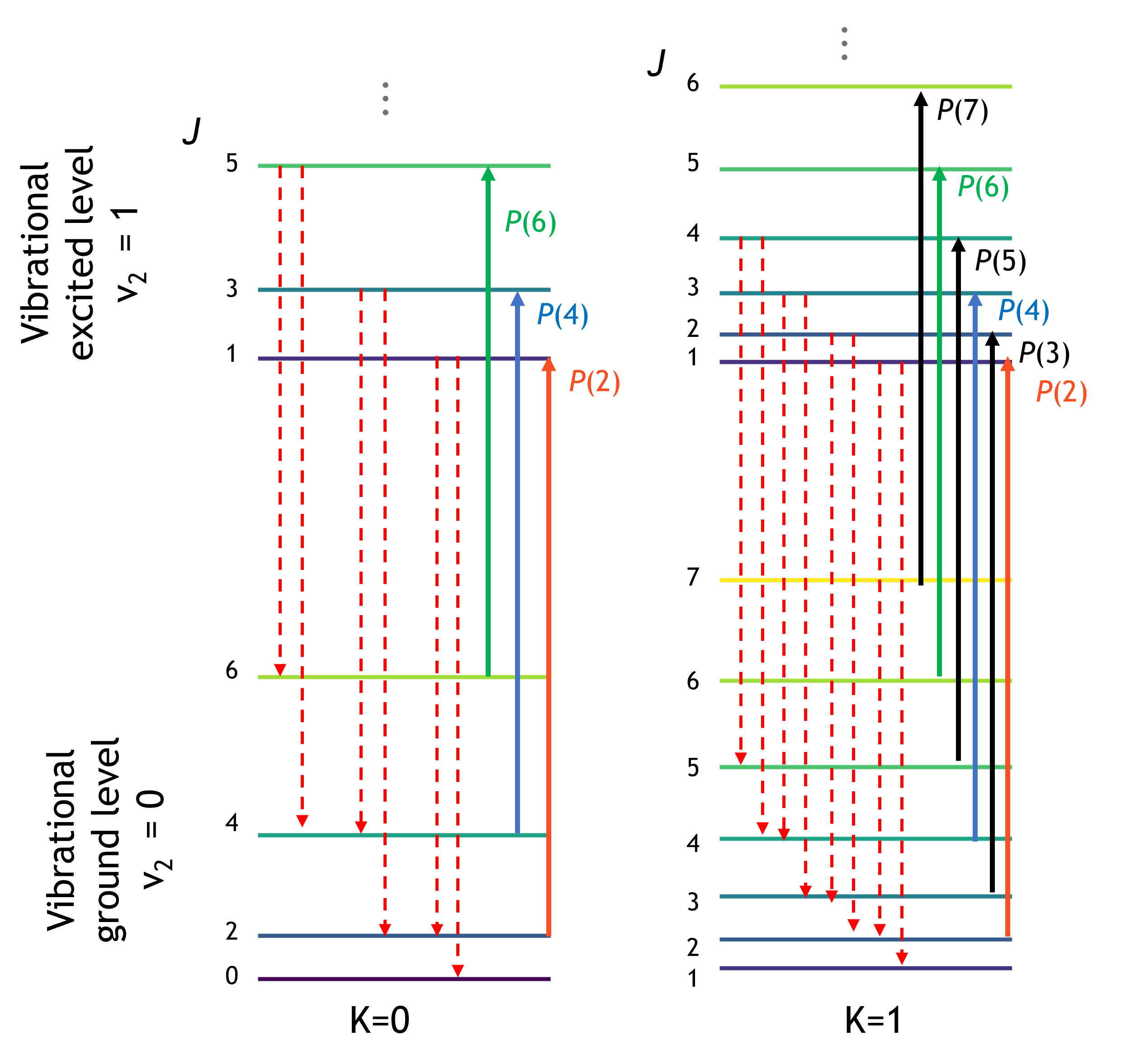}
    \caption{Proposed laser-driven transitions applied for cooling the NH$_3^+$ molecular ion in the $K=0$ (left) and $K=1$ (right) manifolds.
    The $P$ branch transitions
    are denoted with upward arrows.
    The downward cooling transitions are denoted using red dashed lines. While only $P$- and $R$-type decay transitions are possible in the $K=0$ manifold,
    $P$- and $Q$-type decay transitions are accessible in the $K=1$ manifold. $J$ level designations are given beside each state. Laser driven transition with the same color scheme can be pumped using a common laser source e.g. $P(2)$, $P(4)$ and $P(6)$.}
    \label{Cooling_scheme_ppt}
\end{figure}


{
\setlength{\tabcolsep}{4pt}
\renewcommand{\arraystretch}{1.25}
\begin{table*}[htbp]
\caption{\label{tabled:laser pumpings schemes}
Cumulative effect of sequential $P$-branch laser pumping on rotational population redistribution at 300~K.
Additional transitions are added cumulatively within each scheme, and the resulting steady-state populations are shown.
}
\begin{ruledtabular}
\begin{tabular}{l c c |c  c l}
\multicolumn{3}{c|}{NH$_3^+$} & \multicolumn{3}{c}{ND$_3^+$} \\
\hline
Pumped transition & $\nu$ (cm$^{-1}$) &
\multicolumn{2}{c}{Population (\%)} &
$\nu$ (cm$^{-1}$) & Pumped transition \\
\hline
\multicolumn{2}{l}{$K=0$ manifold} & \multicolumn{1}{c|}{$\ket{0,0,0}$} & \multicolumn{1}{c}{$\ket{0,1,0}$} & & \\[2pt]

$\ket{0,2,0}\!\rightarrow\!\ket{1,1,0}$ & 854.5 & 78.9 & 66.9 & 651.1 & $\ket{0,3,0}\!\rightarrow\!\ket{1,2,0}$ \\
$+\ket{0,4,0}\!\rightarrow\!\ket{1,3,0}$ & 807.5 & 89.7 & 84.1 & 629.8 & $+\ket{0,5,0}\!\rightarrow\!\ket{1,4,0}$ \\
$+\ket{0,6,0}\!\rightarrow\!\ket{1,5,0}$ & 756.8 & 89.8 & 85.2 & 608.5 & $+\ket{0,7,0}\!\rightarrow\!\ket{1,6,0}$ \\
$+\ket{0,8,0}\!\rightarrow\!\ket{1,7,0}$ & 702.6 & 89.8 & 85.3 & 587.3 & $+\ket{0,9,0}\!\rightarrow\!\ket{1,8,0}$ \\
$+\ket{0,10,0}\!\rightarrow\!\ket{1,9,0}$ & 644.8 & 89.8 & 85.3 & 566.0 & $+\ket{0,11,0}\!\rightarrow\!\ket{1,10,0}$ \\[4pt]
\hline
\multicolumn{2}{l}{$K=1$ manifold} & \multicolumn{1}{c|}{$\ket{0,1,1}$} & \multicolumn{1}{c}{$\ket{0,1,1}$} & & \\[2pt]
$\ket{0,2,1}\!\rightarrow\!\ket{1,1,1}$ & 855.0 & 49.3 & 19.2 & 661.7 & $\ket{0,2,1}\!\rightarrow\!\ket{1,1,1}$ \\
$+\ket{0,3,1}\!\rightarrow\!\ket{1,2,1}$ & 831.9 & 87.6 & 52.7 & 651.1 & $+\ket{0,3,1}\!\rightarrow\!\ket{1,2,1}$ \\
$+\ket{0,4,1}\!\rightarrow\!\ket{1,3,1}$ & 807.9 & 91.0 & 68.1 & 640.4 & $+\ket{0,4,1}\!\rightarrow\!\ket{1,3,1}$ \\
$+\ket{0,5,1}\!\rightarrow\!\ket{1,4,1}$ & 783.1 & 91.8 & 76.9 & 629.8 & $+\ket{0,5,1}\!\rightarrow\!\ket{1,4,1}$ \\
$+\ket{0,6,1}\!\rightarrow\!\ket{1,5,1}$ & 757.3 & 91.8 & 78.3 & 619.2 & $+\ket{0,6,1}\!\rightarrow\!\ket{1,5,1}$ \\
$+\ket{0,7,1}\!\rightarrow\!\ket{1,6,1}$ & 730.7 & 91.8 & 78.8 & 608.5 & $+\ket{0,7,1}\!\rightarrow\!\ket{1,6,1}$ \\
$+\ket{0,8,1}\!\rightarrow\!\ket{1,7,1}$ & 703.1 & 91.8 & 78.9 & 597.9 & $+\ket{0,8,1}\!\rightarrow\!\ket{1,7,1}$ \\
$+\ket{0,9,1}\!\rightarrow\!\ket{1,8,1}$ & 674.6 & 91.8 & 78.9 & 587.3 & $+\ket{0,9,1}\!\rightarrow\!\ket{1,8,1}$ \\
\end{tabular}
\end{ruledtabular}
\end{table*}
}


In the following analysis, NH$_3^+$ ions are assumed to be initially prepared in the $\ket{0,0,0}$ or $\ket{0,1,1}$ states. Likewise, ND$_3^+$ ions are assumed to be produced in the corresponding rovibrational ground states within the $K=0$ and $K=1$ manifolds, $\ket{0,1,0}$ or $\ket{0,1,1}$.
Such state preparation can be achieved using well-established REMPI experimental schemes, as discussed above. 
Molecular ions produced in the $K=0$ manifold, irrespective of their initial vibrational excitation, tend to relax towards a steady-state distribution confined to the $\ket{\nu=0,\,K=0}$ manifold. Similarly, ions produced in the $K=1$ manifold tend to attain a thermal distribution exclusively within the $\ket{\nu=0,\,K=1}$ manifold. This effective isolation of the $K$ manifolds under BBR-driven dynamics forms the basis for the cooling and state-preparation strategies discussed below.

Targeted optical pumping is applied as represented in FIG. \ref{Cooling_scheme_ppt}. The specific pumping schemes investigated, along with the resulting steady-state population distributions, are summarized in Table~\ref{tabled:laser pumpings schemes}. The cooling mechanism considered here is based on optical pumping of one or more $P$-branch transitions, resulting in a stepwise enhancement of the rovibrational ground-state population. Although the present discussion focuses primarily on the $\nu_2$ vibrational manifold, comparable cooling efficiencies may, in principle, be achieved by driving transitions in the $\nu_4$ manifold. It is important to note, however, that cooling via the $\nu_4$ mode relies on relaxation pathways mediated by the $\nu_2$ manifold.

For the $K=0$ manifold of NH$_3^+$, pumping a single $P$-branch transition, $\ket{0,2,0} \rightarrow \ket{1,1,0}$, enhances the population of the rovibrational ground state $\ket{0,0,0}$ to approximately 79\% at 300~K.
The inclusion of additional $P$-branch transitions leads to a further increase in ground-state population. Our analysis indicates that pumping two $P$-branch transitions at 300~K 
is sufficient to cool the rotational distribution to the rotational ground state within the $K$ manifold ($\ket{0,0,0}$), achieving about 90\% of the total population.
The addition of further $P$ branch transitions achieves a minimal change in the population distribution. 
A similar, albeit slower, trend
is observed for ND$_3^+$ in the $K=0$ manifold. Pumping a single $P$-branch transition, $\ket{0,3,0} \rightarrow \ket{1,2,0}$, increases the population of the rotational ground state ($\ket{0,1,0}$) to approximately 67\%. The inclusion of additional $P$-branch transitions 
results in a gradual enhancement of the ground-state population to $\approx$ 84\%, reaching saturation at around 85\%. 

\begin{figure*}[htbp]
    \centering
    \includegraphics[width=1\linewidth,trim=0 0 0 0,clip]{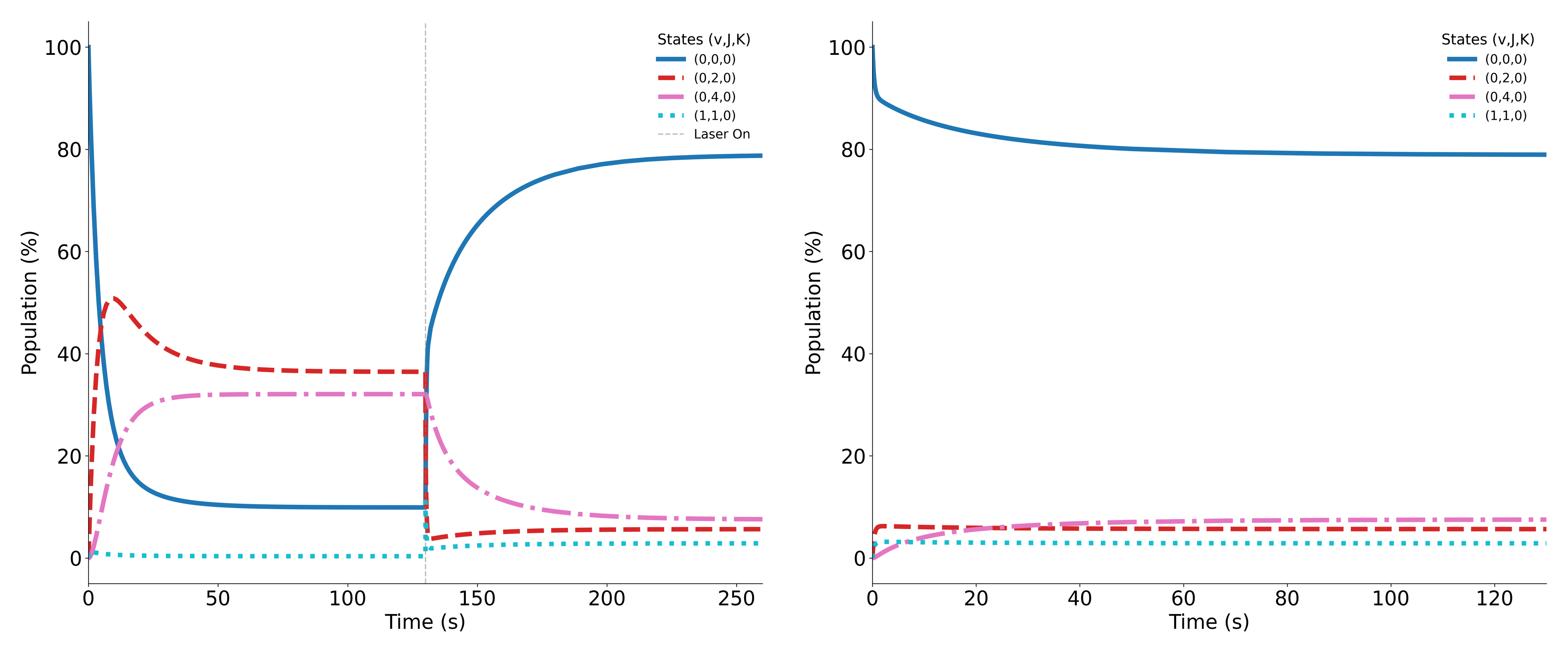}
    \caption{Population dynamics of the $\ket{0,0,0}$, $\ket{0,2,0}$, $\ket{0,4,0}$, and $\ket{1,1,0}$ states of NH$_3^+$ initially prepared in the $\ket{0,0,0}$ state at 300~K. (a) After allowing the ions to equilibrate with the ambient BBR field for $\sim130$~s, a single $P$-branch transition, $\ket{0,2,0}\!\rightarrow\!\ket{1,1,0}$, is applied, leading to efficient population redistribution into the rovibrational ground state, with the $\ket{0,0,0}$ population approaching $\sim80\%$. (b) When the same pumping transition is applied from the outset, the steady-state ground-state population remains unchanged, at $\approx80~\%$; the final population in the rovibrational ground state is the same, independent of the laser switch-on time.}
    \label{laser_decay_dynamics_of_K0_manifold_only}
\end{figure*}

To investigate the role of the pumping-laser switch-on time—specifically, whether population redistribution should first be allowed to equilibrate with the ambient blackbody radiation (BBR) field within a given $K$ manifold, or whether laser pumping should be applied immediately following ion formation—we perform a comparative analysis. Figure~\ref{laser_decay_dynamics_of_K0_manifold_only} summarizes the resulting population dynamics for both the scenarios for the $\ket{0,0,0}$, $\ket{0,2,0}$, $\ket{0,4,0}$, and $\ket{1,1,0}$ states of NH$_3^+$ ions initially prepared state-selectively in the $\ket{0,0,0}$ rovibrational ground state at 300~K.
In Fig.~\ref{laser_decay_dynamics_of_K0_manifold_only}(a), the ions are first allowed to equilibrate with the ambient BBR field for approximately 130~s, during which time BBR-driven redistribution occurs within the $K=0$ manifold. Subsequent application of a single $P$-branch transition, $\ket{0,2,0}\!\rightarrow\!\ket{1,1,0}$, efficiently funnels population into the rovibrational ground state, yielding $\ket{0,0,0}$ populations approaching $\sim80\%$. In 
FIG.~\ref{laser_decay_dynamics_of_K0_manifold_only}(b), 
the dynamics when the same pumping transition is applied from the outset, immediately upon formation of NH$_3^+$ in $\ket{0,0,0}$, is shown. 
Notably, the steady-state population accumulated in $\ket{0,0,0}$ is identical in both cases, demonstrating that the final population is insensitive to the timing of the pumping laser. 
Analogous behavior is observed upon sequential addition of further $P$-branch transitions, namely $\ket{0,4,0}\!\rightarrow\!\ket{1,3,0}$ and $\ket{0,6,0}\!\rightarrow\!\ket{1,5,0}$, achieving a population enhancement of up to $\approx90\%$.
In both scenarios, laser interaction times of approximately 100~s are required to reach steady-state population in $\ket{0,0,0}$. For ND$_3^+$, qualitatively similar behavior is observed; however, owing to its slower internal dynamics, longer pumping durations ($\sim250$ s) are required to achieve population enhancement.

A similar trend is observed for both NH$_3^+$ and ND$_3^+$ molecular ions in the $K=1$ manifold at 300~K. In this case, pumping a single $P$-branch transition ($|0, 2, 1\rangle \rightarrow |1, 1, 1\rangle$) leads to only a modest increase in the population of the $\ket{0,1,1}$ state, reflecting the slower redistribution dynamics within this manifold. For NH$_3^+$, the population rises to approximately 49\% 
after pumping the first transition. The inclusion of additional $P$-branch transitions results in a rapid enhancement, with the ground-state population reaching $\sim92\%$ 
after three additional $P$-branch transitions.
For ND$_3^+$, cooling in the $K=1$ manifold is comparatively less efficient at room temperature, with the population saturating at around 79\% even after multiple transitions are pumped.

These results highlight the combined role of selection rules and isotopic effects in governing the efficiency of rovibrational cooling and further emphasize the strong enhancement achieved under cryogenic 
conditions. In the $K=0$ manifold, $Q$-branch transitions are forbidden, and therefore effective cooling schemes rely exclusively on pumping $P$-branch transitions. In contrast, for the $K=1$ manifold, both $P$- and $Q$-branch transitions are allowed and have been considered for comparison. Exciting additional $Q$-branch transitions did not further improve the results, as detailed in 
Appendix~\ref{app: Q branch cooling}. It is also worth noting the role that nuclear spin statistics play in the redistribution of population. For example,
in the $K=0$ manifold, the set of allowed $P$-branch transitions for ND$_3^+$ is complementary to those of NH$_3^+$ due to the bosonic nuclear-spin statistics of ND$_3^+$, in contrast to the fermionic character of NH$_3^+$. Although even-$J$ rotational states are formally allowed in ND$_3^+$, their statistical weights in the $\nu_2 = 0$ manifold are significantly smaller than those of the odd-$J$ states and can therefore be neglected. 
Overall, the $K=1$ manifold requires a larger number of pumped transitions to reach population levels comparable to those of the $K=0$ manifold. Pumping additional transitions beyond this point yields diminishing returns as the population saturates.

\subsection{Laser-Efficient Cooling Strategies for Mixed $K$-Manifold Populations}

The experimentally relevant scenario in which NH$_3^+$ and ND$_3^+$ molecular ions are produced in a mixed ensemble comprising both $\ket{\nu=0,J,K=0}$ and $\ket{\nu=0,J,K=1}$ states are investigated here. A 1:1 population ratio between the two manifolds is assumed for simplicity, 
although the formalism is readily applicable to arbitrary initial ratios. 
In this case, as summarized in Table~\ref{tabled:laser pumpings schemes}, two pumping transitions are required for the $K=0$ manifold and four transitions for the $K=1$ manifold. Population cannot be transferred from $K=1$ to $K=0$, due to
the $\Delta K = 0$ selection rule.
Under these constraints, the optimal outcome corresponds to population enhancement in the $\ket{0,0,0}$ and $\ket{0,1,1}$ states for NH$_3^+$, and analogously in the $\ket{0,1,0}$ and $\ket{0,1,1}$ states for ND$_3^+$.

A notable feature of the level structure is that transitions in the $K=0$ and $K=1$ manifolds involving the same $J$ quantum numbers occur at nearly identical frequencies. Consequently, a single laser source with modest spectral bandwidth can simultaneously address transitions in both manifolds. This overlap provides the basis for an efficient pumping strategy that minimizes the number of required laser sources while maximizing the achievable ground-state populations.

The laser source is modeled using a Gaussian spectral profile centered at the resonant frequency of the target transition, with a full width at half maximum (FWHM) of 3~cm$^{-1}$, to assess bandwidth effects. Because the relevant transition frequencies are widely separated, multiple Gaussian-profile laser sources are required to efficiently drive the cooling cycle and accumulate population in the rovibrational ground state.

{\renewcommand{\arraystretch}{1.25} 
\begin{table}[htbp]
\caption{\label{tab:optimal pumping schemes}Experimentally viable laser cooling approaches are tabulated along with the corresponding steady-state population in the ro-vibrational ground states. All laser sources are considered to be centered around the stated frequency with a FWHM of 3 cm$^{-1}$.}
\begin{ruledtabular}
\begin{tabular}{l c c}

 & \multicolumn{2}{c}{NH$_3^+$ \% Population at 300~K}\\
\cline{2-3}
Laser freq. (cm$^{-1}$) & $|0, 0, 0\rangle$ & $|0, 1, 1\rangle$\\
\hline

854.5  & 39.4 & 24.6\\
+ 832.0 & 41.6 & 43.7\\
+ 807.7 & 44.8 & 45.5\\
\hline

& \multicolumn{2}{c}{ND$_3^+$ \% Population at 300~K}\\
\cline{2-3}

Laser freq. (cm$^{-1}$) & $|0, 1, 0\rangle$ & $|0, 1, 1\rangle$\\
\hline

661.7  & 7.0 & 9.6\\
+ 651.1 & 33.5 & 26.4\\
+ 640.4 & 33.6 & 34.0\\
+ 629.8 & 42.1 & 38.5\\
\end{tabular}
\end{ruledtabular}
\end{table}
}


Table~\ref{tab:optimal pumping schemes} summarizes the steady-state populations obtained at 300~K through the sequential addition of narrowband laser transitions. For NH$_3^+$, a single pump centered at 854.5~cm$^{-1}$ already produces substantial accumulation in the lowest rovibrational states, yielding $\sim$39\% population in $\ket{0,0,0}$ and $\sim$25\% in $\ket{0,1,1}$. Incorporating additional narrowband transitions progressively enhances redistribution, approaching $\sim$45\% population in each state with three laser frequencies. In contrast, ND$_3^+$ exhibits less efficient population transfer, where a single laser yields only modest accumulation, while successive addition of pumping lasers steadily increases the ground-state populations, reaching $\sim$42\% in $\ket{0,1,0}$ and $\sim$39\% in $\ket{0,1,1}$. Overall, Table~\ref{tab:optimal pumping schemes} clearly demonstrates that achieving comparable rovibrational ground-state populations in ND$_3^+$ requires a significantly larger set of laser sources than for its protic isotopologue.

This complexity arises primarily from the nuclear spin statistics; the statistical weight of the even $\ket{v=0, J=0, K=0}$ states is considerably lower than that of the odd $\ket{0, 1, 0}$ states within the ground vibrational manifold. 
Consequently, the first $P$-branch transition in the $K=0$ manifold emerges as $P(3)$. While a single laser source can be tuned to address this $P(3)$ transition in both the $K=0$ and $K=1$ manifolds simultaneously, it fails to interact with the population accumulated in the lower $J=2$ level of the $K=1$ manifold. To address this "trapped" population within the laser cooling scheme, an additional
laser is required. This source is tasked exclusively with pumping just one 
specific transition, serving no secondary purpose in the cooling cycle. A similar requirement exists for the $\ket{0, 4, 0}$ state. Finally, a fourth laser source is employed as a common pump for both the $\ket{0, 5, 0} \rightarrow \ket{0, 4, 0}$ and $\ket{0, 5, 1} \rightarrow \ket{0, 4, 1}$ transitions, leveraging their spectral proximity.

As the population is distributed across a broader range of $J$ states within the $K=1$ manifold in $\text{ND}_3^+$, omitting a lower-lying transition leads to a significantly reduced population accumulation in the ground state. For the protic isotopologue, a single laser source can pump the equilibrium population into a much narrower distribution, returning up to 40\% of the population to the $\ket{0,0,0}$ starting state. In contrast, for $\text{ND}_3^+$, this efficiency drops to approximately 7\%.

\begin{figure*}[htbp]
    \centering
    \includegraphics[width=0.8\linewidth,trim=0 0 0 0,clip]
    {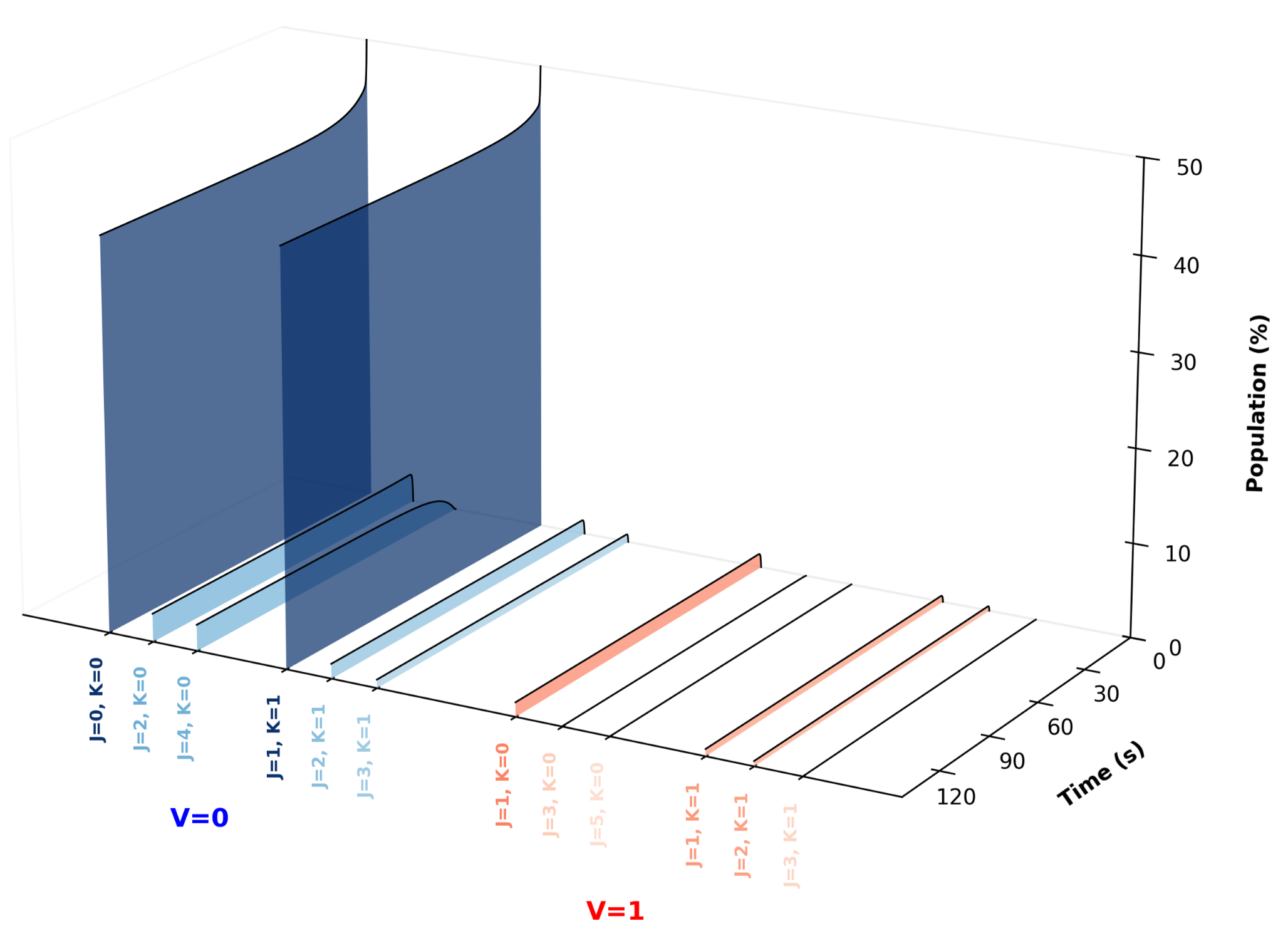}
    \caption{NH$_3^+$ ions are assumed to be produced in an equal ($1{:}1$) mixture of the $\ket{0,0,0}$ and $\ket{0,1,1}$ rovibrational states. Using the laser-pumping scheme developed here, we demonstrate that a common set of narrowband lasers centered at 854.5~cm$^{-1}$ and 832.0~cm$^{-1}$, each with a FWHM of $\sim$3~cm$^{-1}$, can simultaneously address transitions in both the $K=0$ and $K=1$ manifolds.}
    \label{laser_decay_dynamics}
\end{figure*}


In Fig.~\ref{laser_decay_dynamics}, only the first two laser frequencies listed in Table~\ref{tab:optimal pumping schemes} are employed. The first laser simultaneously drives the $P(2)$ transitions in the $K=0$ and $K=1$ manifolds, while the second addresses the $P(3)$ transition in the $K=1$ manifold. Remarkably, this minimal two-laser scheme is already sufficient to produce strong population redistribution, yielding ground-state populations of 42\% in $\ket{0,0,0}$ and 39\% in $\ket{0,1,1}$ for NH$_3^+$.


To assess the impact of laser power variations, it is essential to establish a quantitative connection between the saturation parameter employed in the numerical model and the applied laser irradiance. An analytical expression is derived to determine the optical power threshold required to saturate the targeted transitions. For efficient optical pumping, the stimulated excitation rate must exceed the spontaneous emission rate, characterized by the Einstein $A$ coefficients, by several orders of magnitude. Once the saturation regime is reached, further increases in laser intensity lead to negligible improvements in population transfer efficiency, signaling a plateau in the pumping dynamics. Under these conditions, the rate-limiting step of the cooling cycle shifts from the rapid laser-driven excitation to the substantially slower BBR–induced excitation, which governs the timescale for population redistribution among the rovibrational manifolds. The parameters used to determine the laser spectral density, along with the relevant equations, are derived in the Appendix~\ref{app:laser intensity}.
FIG. ~\ref{laser_saturation} shows 
the range of beam radii and laser power at FWHM of 3 cm$^{-1}$ needed to achieve
saturation.

\begin{figure}[htbp]
    \centering
    \includegraphics[width=1.0\linewidth]{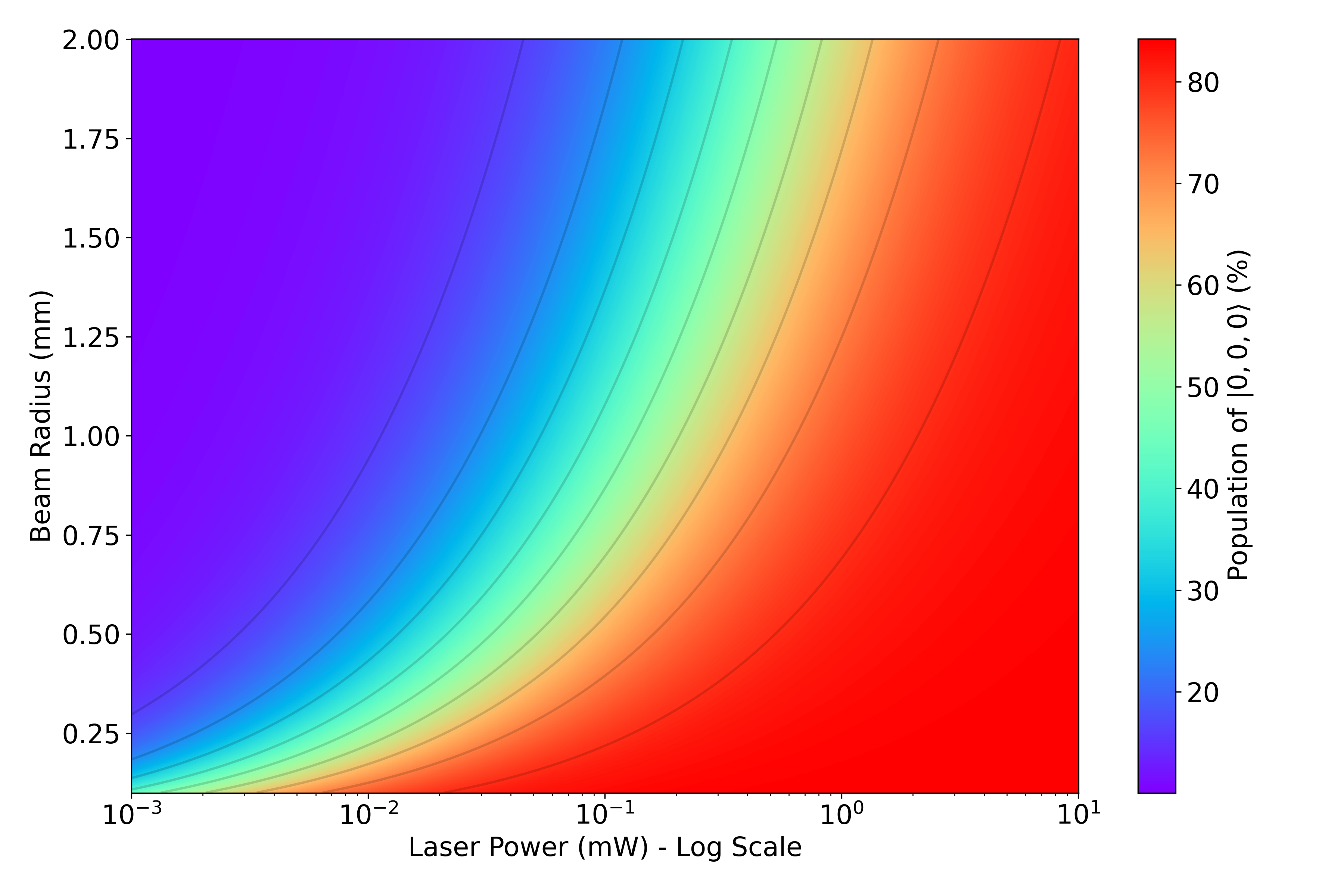}
    \caption{Population of the ground state $\ket{0,0,0}$ of NH$_3^+$ ion as a heatmap over a range of beam radii and laser powers, at an FWHM of 3 cm$^{-1}$.}
    \label{laser_saturation}
\end{figure}

The temperature of the ambient BBR field exerts a critical influence on the cooling kinetics. As the environmental temperature decreases, the thermal population distribution narrows, effectively confining the ions to lower rovibrational levels. While this confinement is theoretically advantageous for high-fidelity state preparation, the suppression of BBR-induced excitation presents a kinetic bottleneck: once thermal redistribution is inhibited, population transfer via optical pumping becomes inefficient. 

The ambient 300~K BBR field facilitates BBR-assisted laser cooling, whereas operation in a cryogenically cooled ion trap enables the effective arrest of a quantum-state–selectively prepared population, provided the ions are produced in the vibrational ground state, allowing the population to be frozen for extended durations.

Fundamentally, the cooling dynamics of NH$_3^+$ and ND$_3^+$ are 
constrained by a $K$-bottleneck, as the dominant $\nu_2$ relaxation pathway strictly follows $\Delta K = 0$ selection rules. While the degenerate $\nu_4$ mode theoretically allows $\Delta K = \pm 1$ transitions, our analysis (detailed in Appendix~\ref{app:nu_4 cooling}) reveals that these transitions are suppressed by rigid Teller selection rules and unfavorable excitation requirements. Due to its lower transition dipole moment and higher saturation intensity—scaling as $I_s \propto \nu^3/|\mu|^2$—the $\nu_4$ mode cannot compete with the faster $\nu_2$ timescales. Consequently, the $\nu_4$ manifold acts only as a slow parallel channel and fails to provide a functional shortcut for inter-$K$ population transfer, leaving the $K$-bottleneck as the defining feature of the timescale for thermalization.


\section{Conclusion}

In this work, we demonstrate that efficient and experimentally realistic rovibrational state preparation of the symmetric-top molecular ions NH$_3^+$ and ND$_3^+$ can be achieved through a judicious combination of laser-driven optical pumping and control of the temperature of the BBR environment. To the best of our knowledge, this study constitutes the first comprehensive, state-resolved simulation of BBR-driven dynamics and laser-assisted cooling pathways in symmetric-top molecular ions. 

Two experimentally viable methodologies naturally emerge from our analysis. In a cryogenically cooled ion trap, quantum-state–selectively prepared populations in the vibrational ground state can be achieved directly
for both isotopologues, enabling interrogation over experimentally optimal timescales with minimal internal-state redistribution. Alternatively, in an ion trap operated at ambient temperatures (300~K), targeted optical pumping schemes can be employed to achieve efficient rotational cooling within a given $K$ manifold. While population transfer between different $K$ manifolds remains forbidden by symmetry-imposed selection rules, our results show that population can nevertheless be accumulated with high efficiency in the lowest rotational ($J$) level of the respective $K$ manifolds for both NH$_3^+$and ND$_3^+$. 
In this regime, since the $\nu_2$ vibrational mode follows a $\Delta K = 0$ selection rule, the symmetric-top ions effectively behave as `pseudo-linear' molecules within a given $K$ manifold. This allows for the adaptation of rotational cooling techniques previously established for linear species like $\text{C}_2\text{H}_2^+$. However, the situation differs from that seen in homonuclear diatomics in that, while the latter lack both rotational and vibrational spectra, these symmetric tops possess a robust vibrational spectrum that can be directly addressed.

Overall, these results establish a clear and practical pathway toward the preparation of internally cold, quantum-state–selected symmetric-top molecular ions in ion traps. The demonstrated strategies significantly relax experimental constraints while delivering high state purity, thereby opening new opportunities for precision spectroscopy, controlled ion–molecule reaction studies, and investigations of quantum-state–resolved chemistry in the cold and ultracold regimes.

\begin{acknowledgments}
The authors gratefully acknowledge Prof. Tim Softley for insightful discussions and constructive suggestions that contributed to this work. ND thanks SERB/ANRF India (CRG/2023/001529) and BRNS India (58/14/21/2023 - BRNS12329) for funding. BRH is grateful to the European Commission (ERC grant no.~948373) and the Leverhulme Trust (grant no.~RPG-2022-264) for funding. AS acknowledges DST-INSPIRE for research fellowship. 
\end{acknowledgments}

\section*{Conflict of Interest}
The authors have no conflicts to disclose.

\section*{Data Availability}
The data that support the findings of this study are available from the corresponding author upon reasonable request.


\appendix
\renewcommand{\thesection}{\Roman{section}}

\section{Efficiency of cooling using $Q$ branch transitions}
\label{app: Q branch cooling}
The evolution of rotational population under laser excitation is governed by the competition between stimulated processes and spontaneous emission. To analyze the direction of population flow, we define $J'$ as the rotational quantum number of the excited vibrational state ($v_2'=1$) and $J''$ as the rotational quantum number of the ground vibrational state ($v_2''=0$). For a parallel transition ($\Delta K = 0$), the spontaneous emission rate (Einstein $A$ coefficient) from an upper state $|v', J', K\rangle$ to a lower state $|v'', J'', K\rangle$ is directly proportional to $A(J' \to J'') \propto \nu^3 \cdot S(J', J'')$, where $S(J', J'')$ are the Hönl-London factors and $\nu$ is the frequency of the energy gap between the corresponding states. The Hönl-London factors for the $P$, $Q$, and $R$ branches (where the suffix denotes the branch of emission from a given upper state $J'$) are given by:
\begin{equation}
\begin{split}
    S_P(J') &= \frac{(J' + 1)^2 - K^2}{J' + 1}, \\
    S_Q(J') &= \frac{(2J' + 1)K^2}{J'(J' + 1)}, \\
    S_R(J') &= \frac{J'^2 - K^2}{J'}
\end{split}
\end{equation}
The $S$ factors are presented in terms of the upper ($J'$) state here. When the laser is tuned to the $Q$-branch ($\Delta J = 0$), a molecule in an initial state $J''_{start}$ is excited to $J' = J''_{start}$. Upon spontaneous decay, the molecule can transition to $J''_{final} = J' + 1$ (increasing $J$) or $J''_{final} = J' - 1$ (decreasing $J$). The ratio of the probabilities for moving "up" versus "down" in the rotational ladder is:
\begin{equation}
    \frac{(J' \to J' + 1)}{(J' \to J' - 1)} \approx \frac{S_P(J')}{S_R(J')} = \frac{J' \left[ (J' + 1)^2 - K^2 \right]}{(J' + 1) \left[ J'^2 - K^2 \right]}
\end{equation}
For the $K=1$ manifold, this expression simplifies to $P_{up}/P_{down} = J'^2(J' + 2) / [(J' - 1)(J' + 1)^2]$. For any $J' > 1$, this ratio is strictly greater than 1 (e.g., for $J'=2$, the ratio is $16/9 \approx 1.78$). Consequently, $Q$-branch excitation acts as a rotational "pump," moving population toward higher $J''$ levels and resulting in heating. Rotational cooling is achieved by exciting the $P$-branch, where $\Delta J = -1$ during absorption, such that the upper state is $J' = J''_{start} - 1$. The selection rules for spontaneous emission allow transitions only to $J''_{final} \in \{J'+1, J', J'- 1\}$. Substituting the laser-excited state $J' = J''_{start} - 1$ into these final states yields $J''_{final} = J''_{start}$, $J''_{final} = J''_{start} - 1$, or $J''_{final} = J''_{start} - 2$. In all permitted decay channels, $J''_{final} \le J''_{start}$. This ensures a unidirectional population flow toward lower rotational levels. Since a $P$-branch transition is physically impossible from the lowest state of the manifold ($J'' = K$ requires $J' = K-1$, which is forbidden), the population accumulates in the $|v''=0, J''=K, K\rangle$ state. This population accumulation effectively cools the rotational distribution to the base of the $K$-stack.

\section{Calculation of Laser Spectral Energy Density}
\label{app:laser intensity}

To model the interaction between laser radiation and molecular species, such as NH$_3^+$, macroscopic laser parameters must be converted into the microscopic spectral energy density, $\rho_{\nu}$. 
This quantity serves as the fundamental input for the Einstein rate equations, which describe the transition probabilities between discrete energy levels via absorption, spontaneous emission, and stimulated emission \cite{einstein1917quantentheorie}. 
Following the nomenclature in Hilborn \cite{hilborn1982einstein}, the spectral irradiance $i(\omega)$ is related to the spectral energy density $\rho(\omega)$ for a directional beam by,
\begin{equation}
   \rho(\omega) = i(\omega)/c, 
\end{equation} 
where $c$ is the speed of light. In terms of frequency $\nu$, this is expressed as

\begin{equation}
    \rho_{\nu} = I_{\nu}/c.
\end{equation}

\noindent While general models often assume power is spread evenly across the beam area ($I_{avg} = P/A$), most lasers exhibit a Gaussian spatial profile \cite{saleh2019fundamentals, loudon2000quantum}. In a Gaussian beam, the intensity is highest at the optical axis and drops toward the edges. To conserve the total energy of the laser, the peak intensity ($I_{peak}$) at the center is exactly twice the average intensity, $I_{peak} = 2P/A$ \cite{demtroder2015laser, metcalf1999laser}. For ions localized in a trap at the geometric center of the beam, they experience this peak intensity rather than the average. Because the rate equations depend linearly on intensity, the choice between $I_{peak}$ and $I_{avg}$ simply accounts for a factor of two in the calculations. Consequently, a laser powerful enough to saturate a transition based on average intensity will certainly be sufficient when the Gaussian peak intensity is considered. To determine the spectral density, the peak intensity is distributed over the laser spectral bandwidth $\Delta\nu$. 
The peak spectral energy density is given by:
\begin{equation}
\rho_{\nu, peak} = \frac{P}{A \cdot c \cdot \Delta\nu}
\end{equation}
This value represents the energy density at the center of the beam and the peak of the laser output. The effective density at a specific molecular transition frequency $\nu_{trans}$ is further modulated by a normalized spectral line shape function $g(\nu)$, such as a Gaussian or Lorentzian profile, to account for the frequency overlap between the laser and the transition\cite{hilborn1982einstein}: 
\begin{equation}
    \rho_{eff} = \rho_{\nu, peak} \cdot g(\nu_{trans}).
\end{equation}  

\noindent By incorporating the spatial and spectral profiles of the laser source directly into the cooling dynamics model, we can accurately simulate the dependence of population pumping on experimental laser power.

\section{Viability of cooling via the $\nu_4$ mode}
\label{app:nu_4 cooling}

Cooling NH$_3^+$ and ND$_3^+$ via the $\nu_2$ vibrational mode is limited by the $\Delta K = 0$ selection rule, which leads to $K$-bottlenecking. In this state, population remains trapped in higher $K$ states even after the rotational levels have been cooled to their lowest possible values within a manifold. While the symmetric bending $\nu_2$ mode is the primary driver for relaxation, a complete model must include the degenerate bending $\nu_4$ manifold. The $\nu_4$ mode has a transition dipole moment aligned perpendicular to the molecular axis. 
The respective Hönl-London factors can be defined as:
\noindent
\begin{equation}
\begin{aligned}
P(J'') &= \frac{(J'' - 1 \pm K'')(J'' \pm K'')}{4J''} \\
Q(J'') &= \frac{(J'' + 1 \mp K'')(J'' \pm K'')(2J'' + 1)}{4J''(J'' + 1)} \\
R(J'') &= \frac{(J'' + 2 \mp K'')(J'' + 1 \mp K'')}{4(J'' + 1)}.
\end{aligned}
\end{equation}
\noindent
where the upper signs refer to transitions with $\Delta K = -1$ and lower to $\Delta K = +1$. The $\nu_4$ mode is theoretically interesting because, as an $E$-symmetry mode, it allows for $\Delta K = \pm 1$ transitions which could potentially facilitate population transfer between $K$-stacks and bypass the bottleneck. In practice, this `shortcut' is suppressed by the coupling of vibrational and rotational angular momenta. Following Teller’s selection rules~\cite{Teller1934}, the $+l$ levels only combine with $\Delta K = +1$ and $-l$ levels with $\Delta K = -1$ transitions, ensuring that population remains trapped within specific $l$ and $K$ combinations. Consequently, the `random walk' across the $K$-manifold is restricted, and the $K$-bottleneck remains a defining feature of the cooling process. Simulations show that $\nu_4$ acts only as a slower parallel relaxation pathway rather than a primary cooling driver, largely because its transition dipole moment is approximately half that of the $\nu_2$ mode. Although the spontaneous emission rate is higher for $\nu_4$ due to its higher frequency, the saturation intensity $I_s$ scales according to: $I_s \propto \frac{\nu^3}{|\mu|^2}$. The combination of a higher frequency and a significantly weaker transition dipole moment means that $\nu_4$ requires a much more intense laser source or tighter focal geometry to achieve saturation. As the $\nu_2$ relaxation timescale is significantly shorter and more efficient it dominates the dynamics, leaving the $\nu_4$ mode as a negligible contributor to the overall thermalization rate of the ions.

\section{Quantitative Assessment of $\nu_4$-Mediated Population Leakage}
\label{app:leakage}

To assess whether blackbody-radiation–induced excitation into the $\nu_4$ vibrational manifold could perturb the cooling cascade, we explicitly compare simulations including only $\nu_2$ redistriibution transitions with those incorporating both $\nu_2$ and $\nu_4$ channels. The rovibrational ground state $\ket{0,0,0}$ does not undergo spontaneous decay but redistributes via BBR-driven stimulated absorption at a rate governed by $B_{abs}\cdot\rho_{BBR}$. The suppression of $\nu_4$ excitation arises from both weaker transition dipole strengths and reduced spectral overlap with the 300 K BBR field. The $\nu_2$ umbrella-bending mode dominates this process: its stimulated absorption coefficients are approximately an order of magnitude larger than those of $\nu_4$ and, when combined with the stronger spectral overlap with the 300 K BBR field, yields effective excitation rates that are 30–40 times faster.

For the $|0,0,0\rangle$ state, the effective BBR excitation rate to $|1,1,0,0\rangle$ (corresponding to $\ket{\nu,J, K, l}$) via $\nu_2$ is $3.15 \times 10^{-1}$ s$^{-1}$, whereas excitation to $|1,1,1,1\rangle$ via $\nu_4$ is only $8.51 \times 10^{-3}$ s$^{-1}$, corresponding to a ratio of approximately 37. A similar hierarchy is observed in the $K=1$ manifold, where $\nu_4$-mediated excitation rates remain more than an order of magnitude smaller than competing $\nu_2$ transitions. The $B_{abs}\cdot\rho_{BBR}$ for selected states in $\nu_2$ and $\nu_4$ are summarised in Table \ref{B-rho_BBR}, with state labels represented as $\ket{\nu,J, K, l}$.

\begin{table}[h]
\centering
\caption{BBR-driven excitation rates from the lowest rotational levels of the $K=0$ and $K=1$ manifolds of NH$_3^+$.}
\begin{tabular}{lc}
\hline
\multicolumn{1}{c}{Transition} &  $B_{abs} \cdot \rho_{BBR}$ (s$^{-1}$) \\ \hline
$\ket{0, 0, 0, 0} \rightarrow |\nu_2=1,1,0,0\rangle$ & $3.15 \times 10^{-1}$ \\
$\ket{0, 0, 0, 0} \rightarrow|\nu_4=1,1,1,1\rangle$  & $8.51 \times 10^{-3}$ \\ 
\hline \hline
$\ket{0,1,1,0} \rightarrow |\nu_2 = 1,1,1,0\rangle$  &  $1.63 \times 10^{-1}$ \\
$\ket{0,1,1,0} \rightarrow|\nu_2 = 1,2,1,0\rangle$  &  $1.53 \times 10^{-1}$ \\
$\ket{0,1,1,0} \rightarrow|\nu_4 = 1,1,0,-1\rangle$ &  $4.40 \times 10^{-3}$ \\
$\ket{0,1,1,0} \rightarrow|\nu_4 = 1,2,2,1\rangle$  &  $8.23 \times 10^{-3}$ \\ \hline
\end{tabular}
\label{B-rho_BBR}

\end{table}

\begin{figure*}[htbp]
    \centering
    \includegraphics[width=1.0\textwidth]{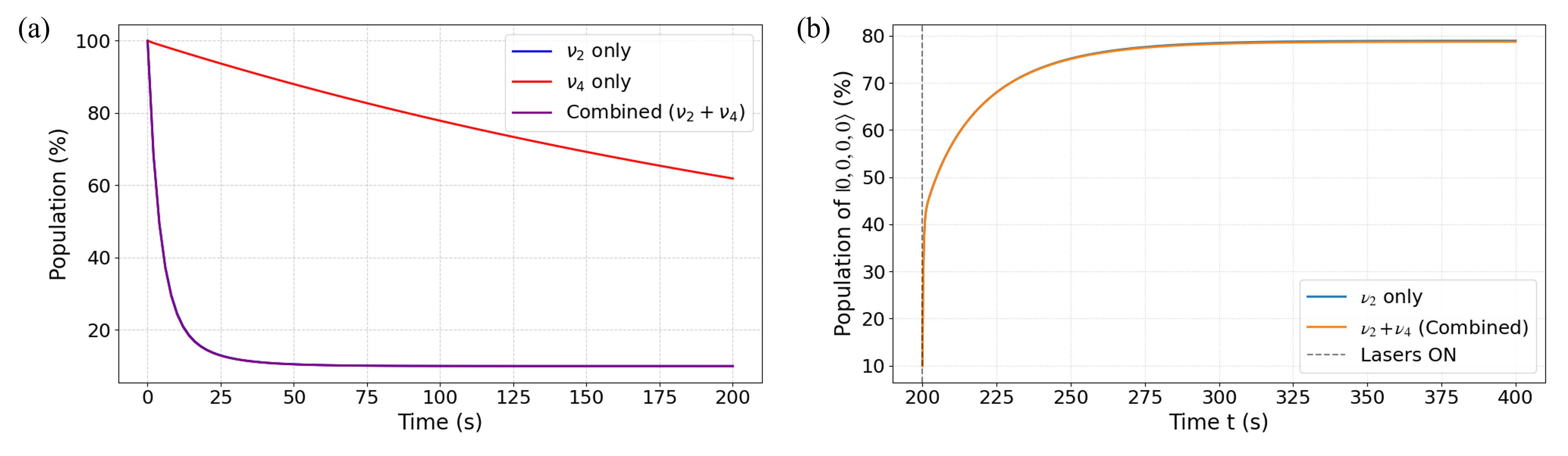} 
    \caption{ (a) Comparison of $|0,0,0\rangle$ population decay considering only $\nu_2$, only $\nu_4$, and both vibrational channels. Note the near-perfect overlap of the $\nu_2$ decay and the combination of $\nu_2$ + $\nu_4$. (b) Comparison of ground-state population accumulation during laser cooling. The plot contrasts the dynamics when considering only $\nu_2$ BBR decay channels versus the inclusion of both $\nu_2$ and $\nu_4$ channels.}
    \label{fig:comparison}
\end{figure*}

FIG~\ref{fig:comparison} (a) shows that the population decay of the $\ket{0,0,0}$ state is virtually indistinguishable when comparing the $\nu_2$-only model with the model including both $\nu_2$ and $\nu_4$ vibrational channels. The final steady-state populations differ by only 0.014\% (9.913\% for $\nu_2$-only versus 9.899\% with $\nu_4$ included). Independent simulations in which decay proceeds exclusively through $\nu_4$ confirm that this pathway is intrinsically slow, with transition rates approximately 30–40 times smaller than those mediated by $\nu_2$. 
The effects of inclusion of $\nu_4$ pathways are further examined in FIG \ref{fig:comparison} (b). Saturating a single $P$-branch transition in the $\nu_2$ manifold yields a ground-state population of 78.951\% after 200~s when only $\nu_2$ decay channels are included. When $\nu_4$-mediated redistribution is incorporated, the population becomes 78.743\%, corresponding to a difference of only 0.21\%.

These negligible deviations confirm that population leakage into the $\nu_4$ manifold does not accumulate to a level capable of altering the cooling cascade. Even over repeated redistribution cycles, the excitation probability into $\nu_4$ remains sufficiently small that cumulative leakage is insignificant. Neglecting this channel in the proposed laser cooling model therefore constitutes a quantitatively justified approximation. Overall, the analysis demonstrates that BBR-driven population flow is decisively governed by the $\nu_2$ mode, with $\nu_4$ providing only a minor perturbative contribution to the redistribution dynamics.


\bibliographystyle{aipnum4-2}

\bibliography{references}

\end{document}